\documentclass[a4paper,11pt]{aa}

\usepackage{graphicx}
\graphicspath{{figures/}}
\usepackage{subcaption}
\usepackage{rotating}
\usepackage[T1]{fontenc}
\usepackage{multirow}
\usepackage[]{algorithm2e}
\usepackage[title]{appendix}
\usepackage{bm}
\usepackage[normalem]{ulem}
\usepackage{txfonts}
\usepackage{hyperref}
\usepackage{xcolor}
\usepackage{bbm}

\usepackage{amsmath}
\allowdisplaybreaks

\usepackage{multirow}
\usepackage{comment}
\usepackage{makecell}
\usepackage{enumitem}
\usepackage{dashrule}
\usepackage{xspace}
\usepackage{booktabs}

\newcommand{\fermi}{{\em Fermi}}
\newcommand{\hess}{{H.E.S.S.}}

\usepackage{comment}

\begin{document}
\title{Learning the cosmic-ray background of Imaging Atmospheric Cherenkov Telescopes with normalizing flows for simulation-based inference}

\author{Christopher Eckner\inst{1, 2}
        \and
        Francesca Calore\inst{3}
        \and
        Jorge Martín Camalich\inst{1, 2}
        \and
        Sami Caroff\inst{4}
        \and
        Miguel Molero Gonzalez\inst{5}}

\institute{
Instituto de Astrofísica de Canarias, E-38205 La Laguna, Tenerife, Spain
\and
Universidad de La Laguna,
Dpto.~Astrofísica, E-38206 La Laguna, Tenerife, Spain\\
\email{ceckner@iac.es, jcamalich@iac.es}
\and
LAPTh, CNRS, F-74000 Annecy, France\\
\email{calore@lapth.cnrs.fr}
\and
LAPP, CNRS, F-74000 Annecy, France\\
\email{caroff@lapp.in2p3.fr}
\and
Centro de Investigaciones Energéticas, Medioambientales y Tecnológicas (CIEMAT), Avda. Complutense 40, 28040 Madrid, Spain\\
\email{miguel.molero@ciemat.es}}

\abstract
{Imaging Atmospheric Cherenkov Telescope (IACT) observations are dominated by a residual cosmic-ray (CR) background. Established estimators require emission-free sky regions. Among such estimators, 3D templates typically rely on discrete binning and in-field refitting without intrinsic systematic uncertainties. Moreover, the lack of an accurate background simulator is a central obstacle to applying 3D simulation-based inference (SBI) to real IACT data.}
{We bring SBI to real IACT data through a unified forward model whose methodological advance combines a continuous, cross-run generative CR background with an out-of-fold calibrated, run-level systematic uncertainty model, enabling neural posterior estimation that marginalizes over background systematics implicitly.}
{Using standard event lists and per-run metadata, we model the residual CR background as an inhomogeneous Poisson point process in the camera frame and reconstructed energy via a rate head and a conditional normalizing flow. Five smooth log-linear deformation modes with calibrated run-to-run covariance capture unrecorded variations. Trained on 55 runs (23.6\,h) of blank fields and masked PKS~2155$-$304 observations from the H.E.S.S.~public data release~1, the generative CR background model enables SBI analyses of the Crab Nebula and the extended source MSH~15$-$52.}
{Cross-fitting reproduces the spatial and spectral background shape to a few per cent with a 15\% run-to-run normalization scatter, and matches five held-out PKS~2155$-$304 runs to per cent accuracy with near-Poisson residuals, whereas unrescaled traditional three-dimensional CR background templates derived from the same data miss the rate by 12\%. On the Crab Nebula and MSH~15$-$52, near the edges of the conditions spanned by the training pool, the CR background's shape is well captured while the rate is offset by 12\% and 10\%, requiring a jointly inferred normalization. The SBI posteriors recover spectrum and morphology in agreement with classical 3D fits (within $\sim\!1\sigma$) and published measurements while propagating calibrated background systematics, establishing a framework ready to scale to larger IACT archives and CTAO.}
{}


\titlerunning{A generative IACT background model for SBI}
\maketitle

\section{Introduction}
\label{sec:intro}

Ground-based gamma-ray astronomy relies on indirect techniques to detect and characterize astrophysical primary high-energy gamma rays because the atmosphere prevents any direct detection from the Earth's surface. One main method is observing the Cherenkov light emitted by secondary relativistic charged particles in the extensive air showers initiated by primary gamma rays in Earth's atmosphere. These instruments, called Imaging Atmospheric Cherenkov Telescopes (IACTs), make the atmosphere an integral part of the detector. They feature large dishes with faceted optical mirrors and a camera into which the collected Cherenkov light is focused. IACTs are typically deployed as an array of multiple instruments to improve the reconstruction of the gamma ray's properties via the stereoscopic observation of the Cherenkov light cone and the resulting shower images taken by the individual telescope cameras \citep{Bose:2021oiq, Holder:GroundbasedGammaRays}. 

The dominant background for IACTs consists of hadronic charged cosmic rays (CRs) whose extensive air showers have been classified as gamma-ray-like. While very efficient background rejection and suppression methods exist (see, e.g., \cite{1985ICRC....3..445H, Albert:2007yd, deNaurois:2009ud, Vuillaume:2021zmv}), it is not possible to fully remove this class of background events. The greatest challenge in ground-based gamma-ray astronomy is to properly estimate or fit the residual CR background in order to analyze an astrophysical gamma-ray signal since the background can be orders of magnitude larger than the expected signal. Since primary charged CRs arrive isotropically, they produce this unavoidable background across the entire Field of View (FoV). However, the observed background rate and spatial distribution depend on the position in the camera frame, telescope pointing, observatory site and atmospheric conditions as well as array configuration, which makes this background an intrinsic part of the Instrument Response Functions (IRFs).

Currently, the most satisfying assessment and treatment of the CR background is data-driven. There are many such approaches, which traditionally rely on observations of the sky that are assumed to be either entirely free of gamma-ray emission or at least contain a region within the FoV that is free of the astrophysical gamma-ray signal to be studied. The earliest technique in the field (to detect the Crab Nebula with the Whipple telescope) was to take ON observations of the target and then to pair it with OFF observations of a nearby empty field so that both runs are taken along the same track in zenith and azimuth \citep{1989ApJ...342..379W}. However, this is a rather inefficient way of spending observation time because it doubles the number of required observations per target where half of them do not even contain the target at all and changing atmospheric or instrument conditions between runs may cause ON and OFF data to differ quite substantially. Therefore, the common contemporary way of taking data is in wobble mode, in which the target is placed at a small offset (typically 0.4$^\circ$--0.7$^\circ$) from the center of the FoV, alternating the direction of the offset between runs \citep{1994APh.....2..137F, Aharonian:2001mz}. The CR background is then estimated simultaneously from the same runs using dedicated techniques: in the reflected-region method, the OFF regions are placed at the same offset from the pointing direction as the ON region, so that they share its acceptance, and the ratio of their solid angles provides the normalization $\alpha$ that enters the significance of \citet{Li:1983fv} and the ON-OFF likelihood of spectral fits \citep{Berge:2006ae}. The ring-background method, used for sky maps, takes the background from an annulus around each test position and corrects for the variation of the acceptance across the annulus with a model of the radial acceptance \citep{Berge:2006ae}.

While these methods are efficient when analyzing individual point-like gamma-ray sources, they have limited applicability elsewhere. This is for two reasons: first, they require emission-free regions in the FoV at a suitable distance from the target; second, some of these methods yield one number for the CR background per energy bin but no spatial structure in the camera FoV\footnote{The ring background method, e.g., does provide spatial information if the ring width is chosen large enough in the OFF region.}. The latter is important when studying diffuse or strongly extended gamma-ray sources in a three-dimensional likelihood-based approach, which fits the spatial and spectral properties of sources jointly \citep{Gammapy:2023gvb}. The necessary three-dimensional CR background templates are derived from large archives of observations with known sources excluded and binned in a certain spatial coordinate system, in energy as well as in observation conditions such as zenith angle and optical efficiency \citep{Mohrmann:2019hfq}. Alternatively, archival runs that match the conditions of each target observation are selected individually or the template is built from the target observations themselves \citep{Wach:2024req, deBonydeLavergne:2026baccmod}. What these methods have in common is that the absolute normalization, and often a spectral tilt, are refitted for each observation in an emission-free part of the FoV. Hence, these approaches reach their limits for sources that fill a large part of the FoV, for diffuse emission and in crowded regions of the Galactic plane, where no emission-free region is available. Moreover, the standard data formats usually carry no further uncertainty for the background model than the freedom of fitting normalization and spectral tilts. Therefore, its systematic uncertainty enters the likelihood only via this additional parametric freedom of the model or it has to be assessed and incorporated with separate methods.

Machine learning has been employed in the context of IACTs for almost two decades \citep[for reviews, see][]{Demichev:2024mnc, Kryukov:2024hka}. Its first and most established application is the separation of gamma rays from hadronic showers to suppress the CR background: random forests have been the standard classifier of MAGIC \citep{Bock:2004td, Albert:2007yd}, and boosted decision trees serve the same purpose for \hess{} and VERITAS \citep{Ohm:2009nw, Krause:2017erl}. With the advent of deep learning, the classification task moved from a handful of image parameters as input to the full camera image. Convolutional and recurrent networks have been trained for event classification and reconstruction on simulations and data of \hess{} \citep{Holch:2017fqm, Shilon:2018xlp}, multi-task architectures have been applied to the data of the first Large-Sized Telescope \citep{Jacquemont:2021cfo, Vuillaume:2021zmv} of the future northern site of the Cherenkov Telescope Array Observatory (CTAO). Recent work extends these approaches to stereoscopic reconstruction for CTAO \citep{Miener:2025ixi}, to graph networks that treat the pixels of several cameras as a single point cloud \citep{Glombitza:2023qen}, to transformers that reconstruct the full event end to end \citep{Jobst:2026fuw}, and to hybrid schemes that combine networks with likelihood fitting \citep{Schwefer:2024bsx} for this matter. Generative networks, first adversarial and more recently diffusion models, have been trained as fast surrogates of the air-shower and image simulation \citep{Elflein:2023jct, Elflein:2026ycz}, and differentiable programming opens the possibility to optimize the design of entire detector arrays end to end \citep{MODE:2022znx}. Almost all of these applications act on the event level in order to process the raw data taken by the instrument. Regarding the high-level analysis, dense feed-forward neural networks employed as deep ensembles have recently been used to separate source and background signals non-parametrically within the observations of a target, relying on a source-free region in the field and on the separability of the spatial and spectral components \citep{Ullmo:2024odq}. 
A model of the CR background that is learned across many independent observations, conditioned on the observation properties and transferable to new fields, has, to our knowledge, not yet been constructed.
Such a versatile neural model of the IACT CR background would be more than just a way to create a background template. If the residual CR rate is learned as a function of the position in the camera, the reconstructed energy and the properties of the observation, it can be evaluated for any observation run without reference to the target field while it also offers the option to be sampled: for a given run, it generates statistically independent realizations of the CR background event list. Together with the IRFs and a model of the gamma-ray sources in a target FoV, it defines a forward simulator of the complete event list of an observation. Such a simulator removes two of the restrictions discussed above. The background no longer has to be measured in emission-free parts of the FoV so that extended sources, diffuse emission and crowded regions can be treated in the same way as isolated point sources. Moreover, the uncertainty of the CR background model can be described by a distribution over its realizations and propagated into the inference avoiding a separate assessment. To profit from these properties, one needs an inference method that works directly with a simulator and does not require an explicit, analytic likelihood function.

Simulation-based inference (SBI) is a Bayesian method in this spirit \citep{Cranmer:2019eaq}. In its neural variants, a network is trained on pairs of parameters and simulated data to approximate either the posterior \citep{Papamakarios:2016ctj}, typically with normalizing flows \citep{Papamakarios:2019fms}, or the ratio of likelihood and evidence \citep{Hermans:2019ioj, Miller:2021hys}. Compared with likelihood-based analyses, this offers several advantages. Any effect that can be simulated, including stochastic processes and instrumental effects, can be included without deriving its contribution to the likelihood. Nuisance parameters are marginalized implicitly, by drawing them from their prior in the simulations, and summary statistics of high-dimensional data can be learned along with the inference. Since the trained networks are amortized, the posterior for a new dataset is obtained at negligible cost, which makes it possible to verify the calibration of the posteriors on large numbers of simulated datasets \citep{Talts:2018zdk, Hermans:2021rqv}. The price to pay is that the inference is only as reliable and robust as the simulator: if the simulated data differ systematically from the observed data, i.e.~model mis-specification, the posteriors can be biased without any indication from the networks themselves \citep{Caron:2022akb}. Successful real data applications of SBI exist in the field of cosmology \citep{Jeffrey:2020xve, Hahn:2022zxa, DES:2024xij}, gravitational wave astronomy \citep{Dax:2021tsq, Dax:2024mcn} and particle physics \citep{ATLAS:2024jry}.

In gamma-ray astronomy, SBI applications have so far focused on the data of the \fermi-LAT, in the context of the Galactic Center excess \citep{Mishra-Sharma:2021oxe, List:2021aer, Christy:2024gsl} and of the faint source populations at high Galactic latitudes \citep{AnauMontel:2022ppb, Eckner:2025waa}. For IACTs, SBI has only been studied on simulated observations at the moment, for instance, to constrain axion-like particles with CTAO \citep{Bhattacharjee:2025ofd}. A central obstacle to applications on real IACT data is that the dominant component of the data, namely the CR background, has no simulator that is accurate enough at the level of individual observations; air-shower simulations of hadronic showers are too expensive and too uncertain for this purpose. 

The main objective of this work is to close this gap by introducing an end-to-end framework that enables SBI on real IACT data, and demonstrating this on the public H.E.S.S.~data release 1 \citep{HESS:2018zix}. Its methodological advance lies in combining three elements into a single data-driven forward model: (i) a continuous, conditional normalizing flow embedded in an inhomogeneous Poisson point process that transfers across independent runs without discrete binning or in-field training; (ii) a low-rank, out-of-fold calibrated systematic handler (a log-Gaussian Cox process) that captures run-to-run instrumental variance natively; and (iii) neural posterior estimation that marginalizes over these calibrated background realizations implicitly rather than profiling per-run normalizations in emission-free control regions. Specifically, we describe the residual CR background of an observation as an inhomogeneous Poisson point process in the camera frame and in reconstructed energy, whose rate and event density are conditioned on the properties of the observation: the zenith and azimuth angles of the pointing and the muon efficiency as a measure of the optical telescope efficiency. The event density is represented by a conditional normalizing flow, the same class of neural density estimators that SBI employs to approximate posteriors. Since the conditioning variables cannot capture every difference between observations, we calibrate the systematic uncertainties on the data and inject them as nuisance into the CR background model. We validate the background model on observations that are not used for its training and analyze two Galactic sources of the release with SBI: the Crab Nebula as a bright point-like reference and the extended pulsar wind nebula MSH~15$-$52, whose morphology and spectrum we infer jointly. We compare the background predictions and the source parameters with those of established methods or literature results, namely reflected-region analyses and background templates built with BAccMod \citep{deBonydeLavergne:2026baccmod}. In this way, we identify where a learned background performs as well as the established methods, where it may not yet and which additional data and metadata would be needed to apply the approach to the archives of current IACTs and to CTAO.

The paper is structured as follows: In Sec.~\ref{sec:flow-bkg}, we provide the technical definitions that make up our generative CR background model and explain how the residual CR intensity is learned and how we include an assessment of the run-to-run systematic uncertainties. Next in Sec.~\ref{sec:framework}, we describe the training and validation framework around the CR background model including a brief outline of the public H.E.S.S.~dataset. This section also states how IACT datasets can be analyzed in the language of SBI and how we employ it in practice for the selected targets. The following Sec.~\ref{sec:results} is dedicated to a report on the various results of the study: First, we quantify the performance of the derived generative CR background model in Sec.~\ref{sec:result-bkg}. Then in Sec.~\ref{sec:result-crab}, we show the outcomes of an SBI analysis employing this model to the Crab Nebula as a target and, lastly, we present the findings of the SBI analysis of MSH 15--52 in Sec.~\ref{sec:result-msh}. The paper closes with a summary and conclusion in Sec.~\ref{sec:conclusion}. The manuscript is supplemented by various appendices (Appendix \ref{app:modes} to \ref{app:msh-infield}) that give more details about the analysis results, systematics and validation checks or alternative approaches to the SBI target analyses. 

\section{Construction of a neural generative background model for Imaging Atmospheric Cherenkov Telescopes}
\label{sec:flow-bkg}

This section outlines the statistical basis and the technical definitions of our model for the residual background of IACTs. By residual background, we refer to the quasi-isotropic events that survive gamma-hadron separation and enter the DL3\footnote{The ``Data Level 3'' (DL3) contains gamma-ray-like events and the IRFs. It is a standardized data format accepted by the full gamma-ray community as the common format as well as the usual input in open source analysis tools like \texttt{gammapy}.} event lists \citep{Deil:2016yrp} as gamma-ray candidates, that is, predominantly misclassified hadronic CRs and CR electrons and positrons alongside any subdominant isotropic extragalactic gamma-ray component. In contrast, structured diffuse gamma-ray emission (such as the Galactic diffuse foreground) is anisotropic on the sky and is treated as part of the astrophysical sky model rather than the instrumental background.
We demand three properties from the CR background model. Given the observation conditions of an arbitrary run, it should \emph{(i)} predict the expected number of background events at any position in the FoV and at any reconstructed energy, \emph{(ii)} generate synthetic realizations of the background for the forward simulator of an SBI analysis, and \emph{(iii)} provide a calibrated statement about its own accuracy. Whereas standard 3D background templates and interpolation schemes \citep[e.g.,][]{Mohrmann:2019hfq, Wach:2024req, deBonydeLavergne:2026baccmod} address only the first requirement (and typically rely on discrete binning and in-field refitting without an intrinsic uncertainty model) combining all three properties in one framework is what makes a data-driven background directly usable inside an SBI forward simulator.

\begin{figure*}[t]
    \centering
    \includegraphics[width=\textwidth]{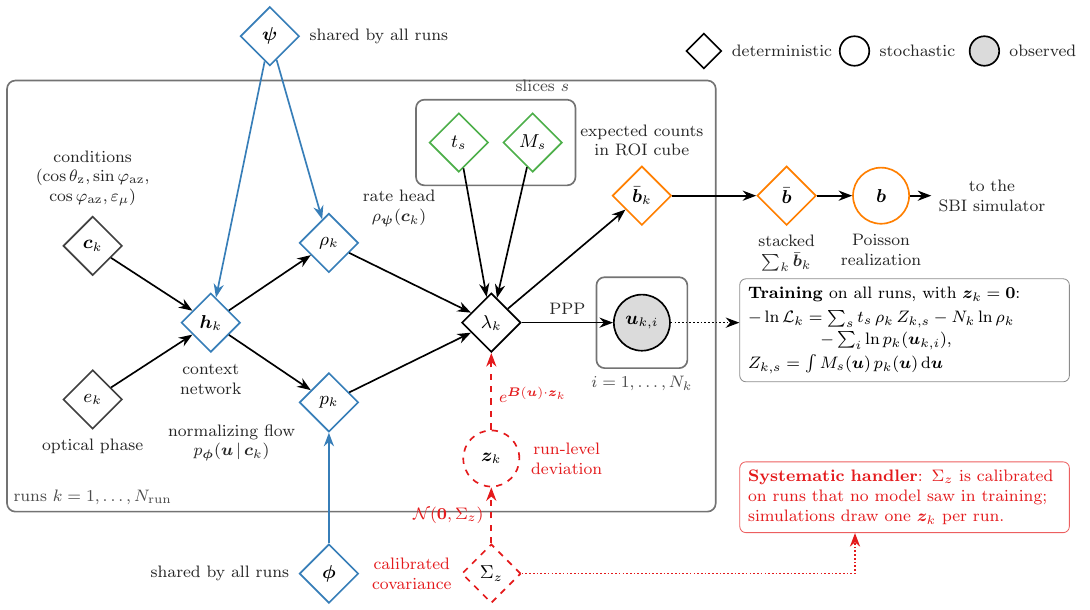}
    \caption{Schematic overview of the generative background model. Diamond-shaped nodes denote deterministic and circular nodes denote stochastic quantities; the shaded node marks the observed events, and plates denote repetition over runs $k$, time slices $s$ and events $i$. A context network turns the observation conditions $\bm c_k$ and the hardware period $e_k$ into a context vector $\bm h_k$, from which a rate head and a conditional normalizing flow compute the rate $\rho_k$ and the shape $p_k$ of the background. Together with the livetime $t_s$ and the mask $M_s$ of each time slice, they define the intensity $\lambda_k$ of the Poisson point process that generates the events. Dashed red elements form the systematic handler: a run-level deviation $\bm z_k\sim\mathcal N(\bm 0,\Sigma_z)$ modulates the intensity through the basis $\bm B(\bm u)$; it is set to zero during training. Orange nodes show how the model enters the forward simulator of an SBI analysis, where the expected counts of all runs are stacked and Poisson-sampled.}
    \label{fig:schematic}
\end{figure*}

We introduce normalizing flows (Sec.~\ref{sec:nf}), the description of the background as an inhomogeneous Poisson point process conditioned on the observation properties (Sec.~\ref{sec:ppp}), the physical effects that shape the background across the FoV (Sec.~\ref{sec:anisotropy}), and a \textit{systematic handler} for the run-to-run deviations that the conditioning cannot capture (Sec.~\ref{sec:sys-handler}). Figure~\ref{fig:schematic} summarizes the model.

\subsection{Normalizing flows}
\label{sec:nf}

A normalizing flow represents a probability density $p(\bm u)$ on $\mathbb{R}^D$ through an invertible and differentiable map $f_{\bm\phi}:\mathbb{R}^D\to\mathbb{R}^D$, which transforms the variable of interest, $\bm u$, into a latent variable $\bm\xi=f_{\bm\phi}(\bm u)$ that follows a simple base density $\pi(\bm\xi)$, usually a standard normal distribution \citep{Rezende:2015ocs, Kobyzev:2019ydm, Papamakarios:2019fms}. The change-of-variables formula yields the density of $\bm u$ exactly,
\begin{equation}
    p_{\bm\phi}(\bm u)=\pi\big(f_{\bm\phi}(\bm u)\big)\left|\det\frac{\partial f_{\bm\phi}}{\partial\bm u}\right|.
    \label{eq:change-of-variables}
\end{equation}
In practice, $f_{\bm\phi}$ is a composition of simple invertible transformations with tractable Jacobian determinants, whose parameters $\bm\phi$ are the weights of neural networks. A flow thus offers two operations at once. It evaluates the normalized density at any point, which allows training by maximum likelihood, and it draws samples by pushing base samples through the inverse map, $\bm u=f^{-1}_{\bm\phi}(\bm\xi)$ with $\bm\xi\sim\pi$. If the networks additionally receive a context vector $\bm h$, the flow represents a conditional density $p_{\bm\phi}(\bm u\,|\,\bm h)$ \citep{Winkler:2019lio}. We use neural spline flows \citep{Durkan:2019nsq}, namely stacks of autoregressive transformations \citep{Germain:2015yft, Papamakarios:2017tec} built from monotonic rational-quadratic splines, which are expressive yet analytically invertible.

Compared with the binned templates that commonly serve as background models in 3D likelihood analyses \citep[e.g.][]{Mohrmann:2019hfq}, a conditional flow offers four advantages. Its density is continuous, so that neither empty bins nor a predefined binning of the FoV, of the energy or of the observation conditions limit the model; it shares statistical strength across the entire space of observation conditions instead of splitting the data into bins; its exact normalization allows us to separate the shape of the background from its rate (Sec.~\ref{sec:ppp}); and sampling is fast, which makes it a natural building block of a forward simulator. Normalizing flows are also density estimators for neural posterior estimation \citep{Cranmer:2019eaq, Papamakarios:2019fms}, which we employ in Sec.~\ref{sec:sbi-analysis}, and they have served to interpolate backgrounds from control regions into a signal region in collider searches \citep{Nachman:2020lpy, Hallin:2021wme}, a task conceptually close to ours.

\subsection{The background as an inhomogeneous Poisson point process}
\label{sec:ppp}

\paragraph{Event coordinates.} We transform the reconstructed arrival direction of every event $i$ of a run $k$ into offsets $(x,y)$ from the instantaneous pointing direction in a tangent-plane FoV frame that is aligned with the altitude-azimuth system, with $y$ pointing towards the zenith, and describe the event by $\bm u=(x,y,\mathcal{E})$ with $\mathcal{E}=\log_{10}(E/\mathrm{TeV})$ and $E$ the reconstructed energy. The dominant structures of the background acceptance are static in this frame, whereas the sky rotates through it during a run \citep{Berge:2006ae, Mohrmann:2019hfq}. The analysis domain $\mathcal{U}_k$ of the run is the FoV disc of radius $R_\mathrm{FoV}$ times its safe energy range.

\paragraph{Point process.} We model the background events of a run as the realization of an inhomogeneous Poisson point process on $\mathcal{U}_k$ \citep{Kingman:1993, Daley:2003}. Such a process is entirely specified by its intensity $\lambda(\bm u)$, the expected number of events per unit livetime and per unit volume of $\bm u$: the number of events in a region $A$ during a livetime $t$ is Poisson distributed with mean $t\int_A\lambda\,\mathrm{d}\bm u$, independently for disjoint regions. This description suits the IACT background. The air showers of CRs arrive independently of each other, and the fraction that survives the gamma-hadron separation is a random thinning of a Poisson process, which is again a Poisson process \citep{Kingman:1993}. Modeling the background thus reduces to learning $\lambda$. We factorize the intensity of run $k$ as
\begin{equation}
    \lambda(\bm u\,|\,\bm c_k)=\rho_{\bm\psi}(\bm c_k)\,p_{\bm\phi}(\bm u\,|\,\bm c_k),
    \qquad
    \int p_{\bm\phi}(\bm u\,|\,\bm c_k)\,\mathrm{d}\bm u=1,
    \label{eq:intensity}
\end{equation}
into a rate $\rho_{\bm\psi}(\bm c_k)$ in units of Hz, which controls \textit{how many} background events a run collects, and the normalized density $p_{\bm\phi}$ of a conditional normalizing flow, which controls \textit{where} in the FoV and \textit{at which energy} they appear.

\paragraph{Conditioning on the observation properties.} The rate as well as the spatial and spectral shape of the background change substantially with the conditions under which a run was taken (Sec.~\ref{sec:anisotropy}). Every background estimate is therefore tied to matching observation conditions, be it through the selection of OFF runs \citep{Berge:2006ae, Wach:2024req} or through the bins of a template model \citep{Mohrmann:2019hfq}. We condition rate and shape on per-run properties that DL3 data provide: the cosine of the zenith angle $\theta_\mathrm{z}$, which traces the depth of the atmosphere; the azimuth angle $\varphi_\mathrm{az}$, encoded by its sine and cosine to respect its periodicity, which together with the zenith angle sets the orientation of the geomagnetic field with respect to the showers; and a measure of the optical throughput such as the muon efficiency $\varepsilon_\mu$ \citep{Chalme-Calvet:2014qxa}. They form the condition vector $\bm c_k$. The hardware period of the run, for instance the optical phase between two re-coatings of the mirrors \citep{Mohrmann:2019hfq, Wach:2024req}, enters as a categorical input $e_k$, so that observational runs of different categories cannot be mixed. Conditioning is what renders the model transferable: the instrumental background is a property of the camera and of the observing conditions rather than of the sky position, so that a model trained on source-free fields predicts the background of any other field observed under matching conditions. Sky-fixed contributions are the exception (Sec.~\ref{sec:anisotropy}).

\paragraph{Likelihood.} Known gamma-ray sources are removed with exclusion regions, which are fixed on the sky and therefore drift through the FoV frame during a run. We divide each run into short time slices $s$, each with livetime $t_s$ and a mask $M_s(\bm u)\in\{0,1\}$ that equals one inside the analysis domain and outside all exclusion regions. In the limit of infinitesimally small cells, the Poisson likelihood of the cell counts of a run becomes the extended unbinned likelihood \citep{Barlow:1990vc}, the continuum counterpart of the Cash statistic of binned analyses \citep{Cash:1979vz}. Inserting Eq.~\eqref{eq:intensity}, the negative log-likelihood of the $N_k$ events of run $k$ reads, up to a model-independent constant,
\begin{align}
    -\ln\mathcal{L}_k(\bm\psi,\bm\phi)={}&\sum_s t_s\,\rho_{\bm\psi}(\bm c_k)\,Z_{k,s}-N_k\ln\rho_{\bm\psi}(\bm c_k)\nonumber\\
    &-\sum_{i=1}^{N_k}\ln p_{\bm\phi}(\bm u_{k,i}\,|\,\bm c_k),
    \label{eq:nll}\\
    Z_{k,s}={}&\int M_s(\bm u)\,p_{\bm\phi}(\bm u\,|\,\bm c_k)\,\mathrm{d}\bm u,
    \label{eq:zks}
\end{align}
where $Z_{k,s}$ is the fraction of the background shape that is visible in slice $s$. The terms separate \textit{how many} from \textit{where}: the first two form the Poisson likelihood of the total number of events and determine the rate, the last is the maximum-likelihood objective of the flow, restricted to the unmasked domain through $Z_{k,s}$.\footnote{Only the product $\rho_{\bm\psi}Z_{k,s}$ and the shape of $p_{\bm\phi}$ inside the analysis domain are constrained by the data, so that $\rho_{\bm\psi}$ is a \textit{nominal} rate; all predictions depend only on the identified combination.} Since the masks cut arbitrary regions out of the domain, we estimate $Z_{k,s}$ by importance sampling with a fixed set of points per run, drawn over the declared safe energy range, which keeps the estimate differentiable with respect to $\bm\phi$. We train the model by minimizing the sum of Eq.~\eqref{eq:nll} over all training runs.

\paragraph{Predictions.} The trained model predicts the expected background counts of run $k$ in any bin $V$ as $\sum_s t_s\int_V M_s\,\lambda\,\mathrm{d}\bm u$. For a region of interest defined on the sky, the sky pixels are mapped into the FoV frame slice by slice, so that the field rotation is accounted for. Summed over the runs of an observation campaign, these predictions form the expected background cube $\bar{\bm b}$, whose Poisson realizations constitute the background component of our forward simulator (Sec.~\ref{sec:sbi-analysis}).

\subsection{Background inhomogeneities and the limits of conditioning}
\label{sec:anisotropy}

The following effects shape the residual background across the FoV. The flow has to learn their imprint, while only some of them are traced by recorded observational properties.
\begin{itemize}[itemindent=0.1in, leftmargin=0.2in]
    \item \textbf{Offset from the pointing direction.} Truncation of images at the camera edge and cuts on the image centroid reduce the acceptance with the offset from the pointing direction, in a strongly energy-dependent way, since energetic showers trigger the array from larger offsets. With a weak hadron rejection at the highest energies, the background may even rise towards the edge of the FoV \citep{Berge:2006ae, Mohrmann:2019hfq}.
    \item \textbf{Zenith angle.} With increasing zenith angle, the energy threshold rises, while above it the effective area grows. The variation of the zenith angle across the FoV turns this into a gradient along the altitude axis, whose sign depends on the energy relative to the threshold \citep{Berge:2006ae, Mohrmann:2019hfq}.
    \item \textbf{Geomagnetic field.} The Lorentz force broadens shower images depending on the angle between the shower axis and the field, so that rate and acceptance depend on the azimuth of the pointing \citep{Chadwick:1999hx, Commichau:2008pe}.
    \item \textbf{Optical throughput.} The degradation of mirrors, light guides and photomultipliers, with abrupt changes at hardware interventions, raises the threshold and lowers the rate \citep{Chalme-Calvet:2014qxa, Wach:2024req}.
    \item \textbf{Atmosphere.} Aerosols, clouds and changes in the density profile alter the Cherenkov light yield and thereby the threshold, the rate and the energy scale \citep{Bernlohr:2000wq, Hahn:2013pfa}; the fitted background normalization of IACT runs correlates with the atmospheric transparency \citep{Mohrmann:2019hfq}. For instance, at La Palma (location of MAGIC and LST-1), an atmospheric haze (Calima) caused by fine sand and dust from the Sahara  can strongly affect data quality or even prevent any data collection. 
    \item \textbf{Night-sky background (NSB) and stars.} The NSB level affects trigger and image-cleaning thresholds, while bright stars disable pixels and create local deficits that are fixed on the sky \citep{Berge:2006ae, MAGIC:2017zph, Archambault:2017jgj}.
    \item \textbf{Instrument state.} Trigger settings, the set of participating telescopes and malfunctioning pixels change the CR rate and acceptance \citep{Wach:2024req}.
\end{itemize}

The conditional intensity absorbs the average imprint of the effects for which a per-run proxy exists. Such proxies can typically be exported as DL3 metadata, while the exact reported quantities will depend on the respective IACT and the available monitoring at the experimental site. Conditioned on the recorded properties, the intensity of a run is therefore itself a random quantity. Even a template model corrected for the atmospheric transparency and the optical phase shows a run-to-run scatter of its fitted normalization of 9--12\% \citep{Mohrmann:2019hfq}, far above the Poisson uncertainty of the event count of a typical run. The background of an ensemble of runs thus forms a doubly stochastic Poisson process, or Cox process \citep{Cox:1955}. A model that ignores the extra variance treats the Poisson noise as the only uncertainty of the background and becomes overconfident precisely for large datasets, where the knowledge of the background limits the sensitivity.

\subsection{The systematic handler}
\label{sec:sys-handler}

\paragraph{Definition.} We describe the deviation of run $k$ from its conditional intensity by a log-linear deformation with a five-dimensional vector of coefficients $\bm z_k$,
\begin{equation}
    \lambda_k(\bm u)=\lambda(\bm u\,|\,\bm c_k)\,\exp\!\big[\bm B(\bm u)^{\mathsf T}\bm z_k\big],
    \qquad
    \bm z_k\sim\mathcal{N}(\bm 0,\Sigma_z),
    \label{eq:sys}
\end{equation}
independently for different runs, with the basis
\begin{equation}
    \bm B(\bm u)=\left(1,\;\frac{r}{R_0}-1,\;\frac{x}{R_0},\;\frac{y}{R_0},\;\frac{\mathcal{E}-\bar{\mathcal{E}}}{\sigma_{\mathcal{E}}}\right)^{\mathsf T},
    \label{eq:basis}
\end{equation}
where $r=(x^2+y^2)^{1/2}$, $R_0=1^\circ$, $\bar{\mathcal{E}}=0$ (i.e.\ 1\,TeV) and $\sigma_{\mathcal{E}}=0.5$. The five modes are a normalization, a radial mode that exchanges events between the center and the edge of the FoV, linear tilts along the azimuthal ($x$) and the altitude ($y$) axis, and a tilt of the reconstructed-energy spectrum; for small coefficients, $z_{k,b}$ is the fractional change induced by mode $b$. Unrecorded changes in the transparency, the throughput or the NSB mainly act on the normalization and the energy tilt, an imperfectly captured zenith or azimuth dependence on the spatial tilts, and errors of the offset-dependent acceptance on the radial mode.

The normalization and the energy tilt correspond to the factor $\Phi\,(E/E_0)^{-\delta}$ with $E_0=1$\,TeV that 3D analyses routinely fit, observation by observation, to an emission-free part of the FoV \citep{Mohrmann:2019hfq, Wach:2024req}, with $\Phi=e^{z_{k,1}}$ and $\delta=-z_{k,5}/(\sigma_{\mathcal{E}}\ln10)$. We do not fit $\bm z_k$ to the target run, which would require an emission-free region in its FoV. Instead, we characterize the population of run-level deviations by their covariance $\Sigma_z$, calibrated once on independent runs, and marginalize over it in the analysis of a target. Eq.~\eqref{eq:sys} thereby turns the background into a log-Gaussian Cox process \citep{Moller:1998} with a covariance function $\bm B(\bm u)^{\mathsf T}\Sigma_z\bm B(\bm u')$ of rank five, the low-dimensional and data-calibrated counterpart of the exponentiated Gaussian random fields with which we modulated the diffuse foreground templates of the \fermi-LAT sky in \citet{Eckner:2025waa}.

\paragraph{Calibration.} For a run $k$ that played no role in training, the log-likelihood as a function of $\bm z$ has, at $\bm z=\bm 0$, the score $\bm S_k-\bm E_k$ and the Fisher information $F_k$, with
\begin{equation}
    \begin{aligned}
        \bm S_k&=\sum_{i=1}^{N_k}\bm B(\bm u_{k,i}),\qquad
        \bm E_k=\sum_st_s\int M_s\,\lambda\,\bm B\,\mathrm{d}\bm u,\\
        F_k&=\sum_st_s\int M_s\,\lambda\,\bm B\bm B^{\mathsf T}\,\mathrm{d}\bm u,
    \end{aligned}
    \label{eq:score}
\end{equation}
A single Newton step estimates the deviation of the run as $\hat{\bm z}_k=F_k^{-1}(\bm S_k-\bm E_k)$. To first order, the covariance of these estimates across runs is the population covariance plus the Poisson noise of each estimate, and moment matching yields
\begin{equation}
    \Sigma_z=\mathrm{Cov}(\hat{\bm z})-\langle F^{-1}\rangle,
    \label{eq:moment-matching}
\end{equation}
projected onto the positive semi-definite matrices. The procedure requires no retraining, but the calibration runs must not have influenced training, preprocessing or model selection, the runs rather than the pixels are the independent units, and $\Sigma_z$ belongs to the predictor it was calibrated for (Sec.~\ref{sec:bkg-val}). The network itself is trained with $\bm z_k=\bm 0$, since the Poisson likelihood yields consistent estimates of the mean intensity even for overdispersed counts \citep{Gourieroux:1984}; the systematic handler supplies the variance that it ignores. In simulations, we draw an independent $\bm z_k$ for every run, so that normalization-like deviations average down in a stack as they should for independent runs, and apply the deformation as $\exp[\bm B^{\mathsf T}\bm z_k-\tfrac12\bm B^{\mathsf T}\Sigma_z\bm B]$.\footnote{Since $\langle\exp(\bm B^{\mathsf T}\bm z)\rangle=\exp(\tfrac12\bm B^{\mathsf T}\Sigma_z\bm B)$, the offset keeps the mean of the simulated intensity equal to the fitted mean intensity.} For target fields whose absolute background rate cannot be predicted, the normalization can be promoted to a parameter that is inferred jointly with the source (Sec.~\ref{sec:sbi-analysis}).

\paragraph{Suitability of the basis and omitted terms.} The basis is small and smooth by design. Every mode extends over the entire FoV or energy range, so that the deformation cannot mimic a compact gamma-ray source; only emission that itself fills a large part of the FoV can become partly degenerate with the normalization. The radial mode and the energy tilt change sign inside the bulk of the events, so that neither acts predominantly as a normalization, and the basis coordinates are physical and model independent, which makes calibrated amplitudes interpretable and comparable across models. We deliberately omit several conceivable terms. \textit{Higher-order and mixed terms}, such as quadratic terms, azimuthal harmonics, a coupling of the radial and the energy dependence or an $xy$ term, would capture finer deviations at the price of a covariance that a few dozen runs cannot constrain; we test the two most plausible of them in Sec.~\ref{sec:bkg-val}. \textit{Localized deficits} caused by bright stars are fixed on the sky and better treated with masks. Finally, \textit{field-coherent deviations} shared by the runs of one target, which would not average down in a stack, and a \textit{condition-dependent mean deviation} $\bar{\bm z}(\bm c)$, which would describe a systematic bias of the model, both require more independent fields than current public data offer; we therefore keep the deviations independent and centered on zero.

\section{Background training and validation strategy}
\label{sec:framework}

We now apply the model of Sec.~\ref{sec:flow-bkg} to public H.E.S.S.\ data. This section describes the data (Sec.~\ref{sec:hess-data}), the training of the model (Sec.~\ref{sec:bkg-train}), the protocols with which we validate it (Sec.~\ref{sec:bkg-val}), and the SBI analyses of gamma-ray sources that build on it (Sec.~\ref{sec:sbi-analysis}). The results follow in Sec.~\ref{sec:results}.

\subsection{The H.E.S.S.~public data release}
\label{sec:hess-data}

The first public data release of H.E.S.S.\ \citep{HESS:2018zix} contains 105 observation runs of $\sim\!28$\,min each, taken between 2004 and 2008 with the four 12-m telescopes of the array in Namibia and reconstructed with the Hillas-based standard cuts. Table~\ref{tab:hess-data} lists the specific contents that we use in this study. For every run, the release provides the event list with arrival direction, reconstructed energy and time, the good time intervals, and full-enclosure instrument response functions (IRFs), i.e.~effective area, energy dispersion and point spread function (PSF), each tabulated in the offset from the pointing direction. It does not include a background IRF. The observation index supplies the per-run metadata from which we build the conditions of Sec.~\ref{sec:ppp}: the pointing, for which we compute the zenith and azimuth angles at the middle of time of the run; the muon efficiency $\varepsilon_\mu$ as a proxy of the optical throughput; and the safe energy range, whose lower edge $E_\mathrm{safe}$ (the energy at which the energy bias equals 10\% at an offset of $1^\circ$ \citep{HESS:2018zix}) lies between 0.2 and 1.3\,TeV depending on the zenith angle. The optical phase follows from the date of the run. For H.E.S.S., these are periods of stable optical efficiency, which are separated by specific events that drastically change the efficiency, e.g.,  mirror re-coating or the cleaning of the Winston cones (light guides attached to the camera pixels). All runs for this study belong to phase 1/1b \cite{Wach:2024req}. The public H.E.S.S.~dataset does contain more observations of PKS~2155$-$304 from 2008, which were however taken in the optical phase 1c. We decided to exclude runs of phase 1c from our analysis to avoid mixing with a phase that has very little data coverage. In principle, it would be interesting to investigate if a generative CR background model can be derived that transfers among many different optical phases; to this end, more training data would be necessary than the public dataset contains. Throughout this work, we use a FoV radius $R_\mathrm{FoV}=2^\circ$ and an upper energy limit of 100\,TeV.

\begin{table*}[t]
\centering
\small
\setlength{\tabcolsep}{4pt}
\begin{tabular}{lrrccc p{5.2cm}}
\toprule
Dataset & runs & livetime [h] & zenith [deg] & azimuth [deg] & $\varepsilon_\mu$ & Role in this work \\
\midrule
Off (blank fields)            & 45 & 19.2 & 1.6--52.8  & 0--360 (none in 258--314) & 0.853--1.189 & training pool of the background model \\
PKS 2155$-$304 flare          & 10 &  4.4 & 7.5--50.3  & 110--251 & 1.010--1.030 & training pool, blazar excluded within $0.5^\circ$ \\
PKS 2155$-$304 flare          &  5 &  2.2 & 8.8--43.2  & 110--250 & 1.013--1.023 & held out: validation of the background model \\
Crab Nebula                   &  4 &  1.8 & 45.4--48.0 & 4--23    & 0.907--0.920 & SBI target: point source, spectrum \\
MSH 15$-$52                   & 20 &  8.3 & 35.5--40.6 & 168--199 & 1.145--1.221 & SBI target: extended source, morphology and spectrum \\
\bottomrule
\end{tabular}
\caption{Content of the H.E.S.S.~public data release used in this work. Zenith and azimuth angles (azimuth from north towards east) refer to the pointing direction at the middle of each run, $\varepsilon_\mu$ is the muon efficiency. The training pool of Sec.~\ref{sec:bkg-train} comprises the off runs and ten flare runs of PKS~2155$-$304; the other five flare runs are held out for validation (Sec.~\ref{sec:bkg-val}). The remaining fields are the targets of the SBI analyses (Sec.~\ref{sec:result-crab} and Sec.~\ref{sec:result-msh}).}
\label{tab:hess-data}
\end{table*}

\subsection{Training the generative model}
\label{sec:bkg-train}

\paragraph{Training pool.} The model learns from 55 runs: the 45 off runs, which point at fields without known gamma-ray emission, and ten runs of the 2006 flare of PKS~2155$-$304, from which we exclude a disc of radius $0.5^\circ$ around the blazar. We chose this radius such that the leakage of source photons beyond it, estimated from the PSF, stays below 1\% of the local background in every energy band. Since the blazar is a point source at high Galactic latitude, the remaining FoV then samples the CR background in the same way as an off run. The other five flare runs, selected before any analysis to cover the zenith range of the flare, are held out for validation (Sec.~\ref{sec:bkg-val}). The pool amounts to a livetime of 23.6\,h and spans zenith angles of 1.6--52.8$^\circ$ and muon efficiencies of 0.853--1.189. We divide each run into time slices of at most 120\,s and at most $0.5^\circ$ of field rotation, with a median of 20 slices per run, and project the exclusion regions into the FoV at the middle of each slice. We train the model on all available observation conditions (zenith, azimuth, $\varepsilon_{\mu}$) without explicitly checking, which of them has the highest impact on the prediction accuracy of the model. A dedicated analysis along these lines will be the subject of future work based on a larger pool of training runs and observation conditions. 

\paragraph{Architecture and optimization.} The context network maps the standardized condition vector, concatenated with a learned four-dimensional embedding of the optical phase, through two fully connected layers of 64 units with GELU activations \citep{Hendrycks:2016qxa} onto the context $\bm h_k$. The rate head is a linear map from $\bm h_k$ to $\ln\rho$. The flow is a neural spline flow of six autoregressive transformations with eight spline bins each, whose masked networks have two hidden layers of 128 units and receive $\bm h_k$ as context. The model has 209\,507 trainable parameters. Event coordinates and conditions are standardized with the mean and the standard deviation of the training runs only. The importance sample for $Z_{k,s}$ contains 8192 points per run, drawn uniformly over the FoV disc and, in energy, from a smoothed histogram of the events of the run within its declared safe range, mixed with a uniform component of 10\% weight. We minimize the sum of Eq.~\eqref{eq:nll} over the training runs with the Adam optimizer \citep{Kingma:2014vow} for a fixed budget of 400 steps and with a learning rate of $5\times10^{-4}$ that decays on a cosine schedule. Each step draws 24 runs and, for each of them, random subsamples of 512 events and 512 importance points, which yield an unbiased estimate of the likelihood. We keep the final state of the network and employ no early stopping condition. The implementation uses zuko \citep{Rozet:2022zuko} and PyTorch \citep{Paszke:2019xhz}.

\subsection{Validation protocols for the generative model}
\label{sec:bkg-val}

\paragraph{Cross-fitting and calibration of the systematic handler.} A model reproduces its training runs better than new ones. The tests of the model as well as the calibration of $\Sigma_z$ therefore require, for every run, a prediction from models that did not see this run. With only 55 runs, a permanent split into training and test sets would waste most of the pool. We therefore cross-fit: we divide the pool into five groups of 11 runs, stratified in zenith angle and muon efficiency, and train five models, each on the 44 runs outside one group. Every run then receives a prediction from a model that did not see it but was trained on almost the full pool. Because the outcome depends on the particular grouping, we repeat this procedure for five independent groupings, i.e.\ with 25 models in total. For every pool run, the five models that excluded it, one from each grouping, form an ensemble whose averaged intensity serves as the out-of-fold prediction. From the deviations $\hat{\bm z}_k$ of these predictions (Eq.~\eqref{eq:score}), we obtain a single $\Sigma_z$ via Eq.~\eqref{eq:moment-matching}. For fields outside the pool, we average the intensities of the five models of one grouping, which together have seen all 55 runs, and refer to this ensemble as the \emph{benchmark generative background model}. Both ensembles consist of five members trained on 44 runs each, so that $\Sigma_z$ is calibrated for the kind of predictor that we apply to new fields. The fixed training budget ensures that the held-out runs enter neither the training nor the model selection.

\paragraph{Background bias.} We summarize the agreement between prediction and data for a set of runs by the bias $\sum_k N_k^\mathrm{obs}/\sum_k N_k^\mathrm{pred}-1$ of the stacked counts between 0.5 and 20\,TeV, inside the FoV disc and outside the exclusion regions. For every run, the lower energy limit is the larger of 0.5\,TeV and its $E_\mathrm{safe}$, identically for data and prediction. We fixed the lower energy threshold at 0.5 TeV as a compromise to accommodate all considered fields with one number; the Crab observations have particularly high $E_{\rm safe}$ between 0.66--0.87 TeV. The bias estimation must be with respect to entire runs since $\Sigma_z$ is designed to describe a common deviation of all events of a run. We therefore derive all uncertainties by bootstrapping whole runs.

\paragraph{Significance maps.} To test the spatial structure, we compare the observed counts $N$ with the predicted counts $B$ within a top-hat circle of radius $0.06^\circ$ around every pixel of size $0.02^\circ$ and compute the significance of \citet{Li:1983fv} in the limit of a known background,
\begin{equation}
    S=\mathrm{sign}(N-B)\sqrt{2\left[N\ln\frac{N}{B}-(N-B)\right]}.
    \label{eq:lima}
\end{equation}
For a correct background model, $S$ follows a standard normal distribution in source-free regions. The width of this distribution probes spatial structure beyond Poisson fluctuations, whereas its mean mainly reflects the normalization of the prediction, scaled by the square root of the counts per circle; the mean therefore depends on exposure, circle radius and energy band. Since neighboring pixels share events, their significances are correlated, and the width of the pixel distribution does not translate into an uncertainty of its mean. We also employ bootstrapping of whole runs here.

\paragraph{Held-out runs and transfer.} The five held-out flare runs of PKS~2155$-$304 test the benchmark generative background model on runs of a field and campaign that are represented in training; they probe interpolation, not the transfer to entirely new fields in the sky. For these runs, we compare the observed stacked counts with their predictive distribution including the variance from $\Sigma_z$. The targets of Sec.~\ref{sec:sbi-analysis} test the transfer to fields absent from training. Before any transfer, we check how densely the training runs cover the conditions of the target runs. Finally, we tested whether the five-mode systematic basis misses a coupling between the radial and the energy dependence of the acceptance, or an $xy$ term. In short, we do not find evidence that such terms are required at the sensitivity of the H.E.S.S.~public data (see Appendix~\ref{app:modes} for more details).

\subsection{Gamma-ray analyses in the language of simulation-based inference}
\label{sec:sbi-analysis}

\paragraph{Targets.} The release offers two test cases of increasing difficulty (Table~\ref{tab:hess-data}): the Crab Nebula, a point-like source observed at the upper end of the zenith range of the training pool; and MSH~15$-$52, an elongated source with a semi-major axis of about $0.2^\circ$, whose runs lie at muon efficiencies above those of most training runs.

\paragraph{Setup of the forward model.} We rely on the functionalities of the gamma-ray simulation and analysis software \texttt{gammapy} \citep{Gammapy:2023gvb} to set up our forward simulator of gamma-ray emission per target. For each target, we simulate a three-dimensional counts cube with $4^\circ\times4^\circ$ at $0.04^\circ$ pixels and eight bins per decade of reconstructed energy (energy range depends on the specific target) centered on the position of the respective target as reported in the literature \citep{Mohrmann:2019hfq}. The analysis band starts at the first energy bin edge above the largest $E_\mathrm{safe}$ of the runs, so that no run contributes below its threshold. \texttt{gammapy} internally reduces the runs with their full-enclosure IRFs to a stacked exposure cube with 20 bins of true energy, and to the energy dispersion and the PSF at the source position. The expected counts are then consistently produced by specifying a certain gamma-ray spectral shape $\mathrm{d}N/\mathrm{d}E$ and a normalized spatial template $T$ for the target. 

The CR background enters as the expected cube $\bar{\bm b}$ of the benchmark generative background model (Sec.~\ref{sec:ppp}). Exclusion regions of neighboring sources are applied identically to the observed cube, the background prediction, the source response and every simulated cube. In every simulation, we draw a deviation $\bm z_k$ per run from $\Sigma_z$, so that the background varies as calibrated on independent runs. Where the absolute background rate of a target cannot be predicted  (i.e.~whenever the systematic uncertainty of the generative background's normalization is smaller than the relative difference between the model's mean predicted CR rate and the observed one), we additionally scale the background by a normalization $\eta$, which we infer jointly with the source parameters under a uniform prior on $\log_{10}\eta$ whose range the posterior must not reach. A Poisson draw of the summed expectation yields the simulated cube $\bm x$.

\paragraph{Neural posterior estimation.} We approximate the posterior $p(\bm\theta\,|\,\bm x)$ of the source parameters $\bm\theta$, including $\eta$ where applicable, by a conditional neural spline flow $q(\bm\theta\,|\,\bm s(\bm x))$. It is trained jointly with a summary network $\bm s$ by minimizing $-\ln q(\bm\theta\,|\,\bm s(\bm x))$ over pairs $(\bm\theta,\bm x)$ drawn from the prior and the simulator \citep{Papamakarios:2016ctj, Cranmer:2019eaq}. This approach is commonly called \emph{neural posterior estimation} in SBI. The summary network is a convolutional network that treats the energy bins of the cube as channels. The counts are first summed in blocks of pixels, variance-stabilized by a square-root transform and standardized voxel by voxel with the moments of the training simulations. A fixed depthwise convolution with the PSF of every energy channel acts as a matched filter, and trainable convolutional layers, adaptive pooling and a fully connected layer produce the summary. We train on $\mathcal{O}(10^4)$ simulations (specified later per target in Sec.~\ref{sec:results}), of which 20\% serve for validation, redraw the Poisson noise of each simulation in every epoch, and keep the epoch with the lowest validation loss. The implementation uses the software packages \texttt{swyft} \citep{Miller:2021hys} and \texttt{zuko}. 

\paragraph{Validation of the posteriors.} A posterior is only useful if it is calibrated and faithful, that is, if the observed data are a plausible outcome of the simulator. Appendix~\ref{app:sbi-val} defines the diagnostics we summarize in what follows. We test calibration with simulation-based calibration \citep{Talts:2018zdk}, i.e.\ the uniformity of the rank of the true parameter among posterior draws for 1000 fresh simulations as well as with the expected coverage of joint highest-posterior-density regions \citep{Hermans:2021rqv}. Posterior predictive checks compare a statistic of the observed cube, namely the total counts, with simulations at posterior draws. Since a predictive distribution that is much wider or much narrower than the observed variation passes or fails such a check trivially, we report the ratio of predictive standard deviation to observed value alongside every $p$-value. As a reference, we perform a standard template-based fit with the same forward model to the observed cube by maximizing the Poisson likelihood and a free background normalization.

\section{Results and interpretation}
\label{sec:results}

We report here the results of the analysis and validation strategies outlined in the previous section. First, we characterize and quantify the performance of the generative CR background model in Sec.~\ref{sec:result-bkg}. Then, we study our first astrophysical target, the Crab Nebula, with an SBI analysis leveraging this background model in Sec.~\ref{sec:result-crab}. Lastly, we turn towards the extended source MSH 15--52 in Sec.~\ref{sec:result-msh} and employ a more elaborate SBI analysis to investigate its spatial and spectral features.

\subsection{Performance of the generative background model}
\label{sec:result-bkg}

\paragraph{Uncertainty of the prediction.} The cross-fitted calibration attributes almost all of the run-to-run variation to the normalization mode, with an amplitude of 15\% (95\% run-bootstrap interval 12--18\%), whereas the radial mode, the two tilts and the energy tilt remain between 1.4\% and 3.8\%. The model thus predicts the spatial and spectral shape of the background of an unseen run to a few per cent, but its normalization only to about 15\%, several times the Poisson uncertainty of a typical run. These deviations average down in a stack of runs: on the 55 pool runs, the stacked cross-fitted prediction deviates from the observed counts by $+0.1\pm1.9\%$.

\begin{figure*}[t]
    \centering
    \begin{subfigure}{0.32\textwidth}
        \includegraphics[width=\textwidth]{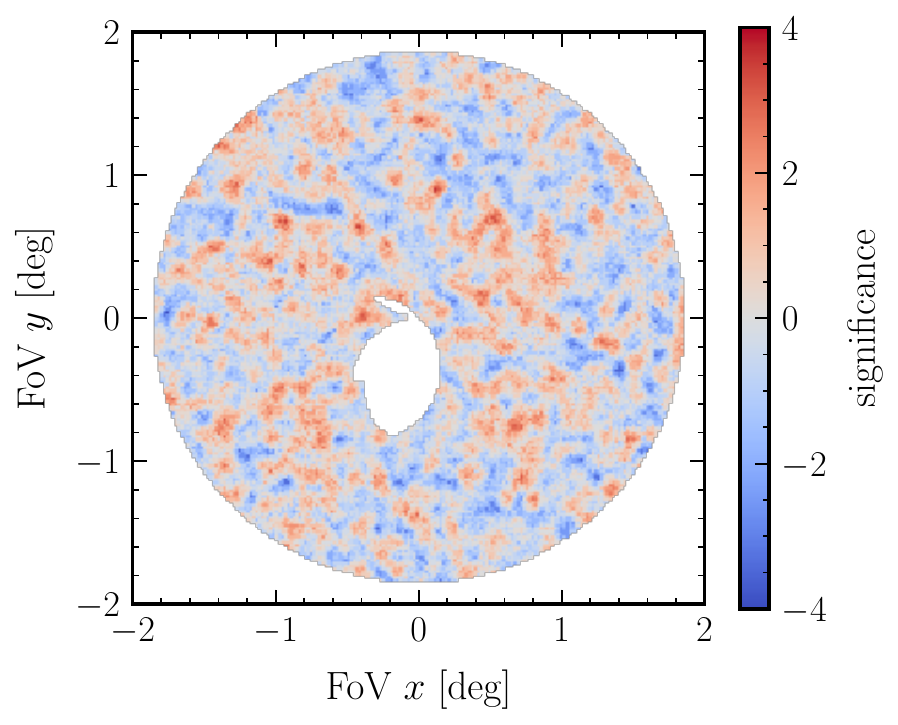}
        \caption{Generative model}
    \end{subfigure}
    \hfill
    \begin{subfigure}{0.32\textwidth}
        \includegraphics[width=\textwidth]{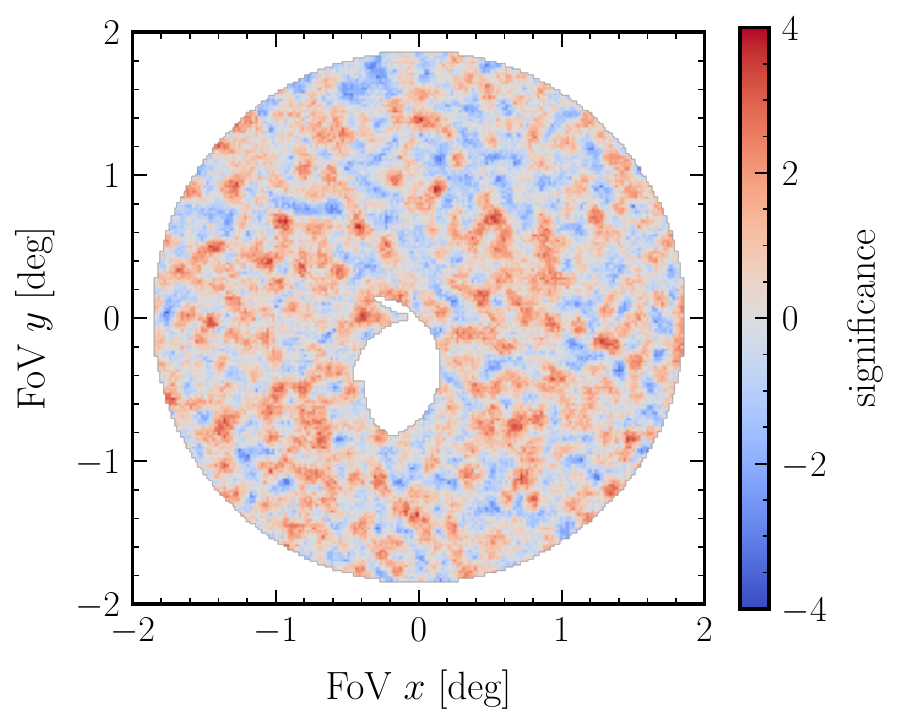}
        \caption{BAccMod}
    \end{subfigure}
    \hfill
    \begin{subfigure}{0.32\textwidth}
        \includegraphics[width=\textwidth]{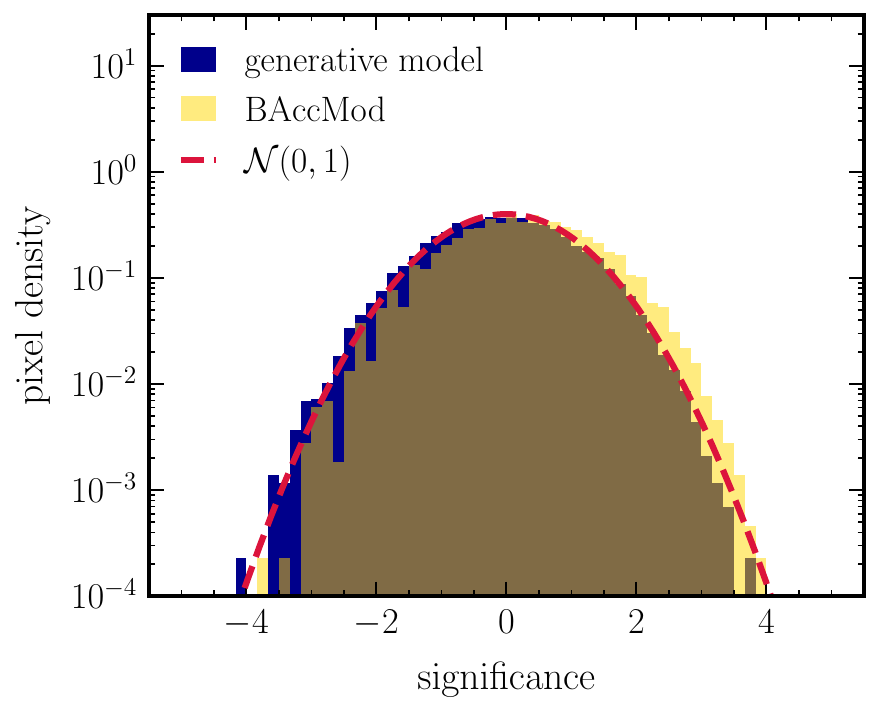}
        \caption{Pixel distributions}
    \end{subfigure}
    \caption{Significance maps (Eq.~\ref{eq:lima}) of the five held-out flare runs of PKS~2155$-$304, stacked in the FoV frame between $\max(0.5\,\mathrm{TeV},E_\mathrm{safe})$ and 20\,TeV, for the prediction of the benchmark generative background model (a) and of the best BAccMod setup built from the same 55 runs (b). In every run, the blazar lies $0.5^\circ$ from the FoV center but in different directions, so that each exclusion disc of $0.5^\circ$ radius touches the center; pixels excluded in at least three of the five runs as well as pixels beyond $1.85^\circ$, are blank. Panel (c) compares the pixel distributions with a standard normal distribution: $\mu=-0.07$, $\sigma=1.06$ for the generative model (blue) and $\mu=+0.27$, $\sigma=1.07$ for BAccMod (yellow).}
    \label{fig:holdout}
\end{figure*}

\paragraph{Held-out runs.} Figure~\ref{fig:holdout} shows the significance maps of the stacked five held-out flare runs of PKS~2155$-$304 for the prediction of the benchmark generative background model (lower energy threshold at 0.5 TeV). The stacked prediction matches the 9085 observed events to 0.1\%, well within the predictive uncertainty of 7\% that the systematic handler assigns to these runs. The significances follow a standard normal distribution with a mean of $-0.07$ and a width of $\sigma=1.06$, close to the values of $-0.05$ and 1.01 that Poisson fluctuations alone produce for this stack. We take this width as the benchmark of the background model for the target fields below. We checked that lowering the energy threshold for this validation check to 0.25 TeV (lowest $E_{\rm safe}$ among the held-out runs) produces almost identical performance results.

\paragraph{Comparison with BAccMod.} Among the BAccMod setups we tested (Appendix~\ref{app:baccmod}), the radial model built, for every run, from the pool runs within $\pm5^\circ$ of its zenith angle performs best. Built from the same 55 runs, it reproduces the spatial structure of the held-out runs equally well ($\sigma=1.07$) but underestimates their background by 12\% (mean significance $+0.27$). Rescaling the prediction of every run to its own counts removes this offset, but requires an emission-free part of the FoV. The absolute prediction, on which the SBI analyses rely, is thus where the conditioning of the generative model pays off. In fact, not only does continuous conditioning in the generative CR background model improve the unrescaled prediction on the held-out PKS runs, it simultaneously supplies a calibrated run-to-run uncertainty (7\% for the stack) that binned templates do not bring natively.

The cross-fitted comparison over the whole pool, the other BAccMod setups, and a weak radial residual that both predictions share are reported in Appendices~\ref{app:modes} and~\ref{app:baccmod}.

\subsection{The Crab Nebula with SBI}
\label{sec:result-crab}

\begin{figure*}[t]
    \centering
    \begin{subfigure}{0.35\textwidth}
        \includegraphics[width=\textwidth]{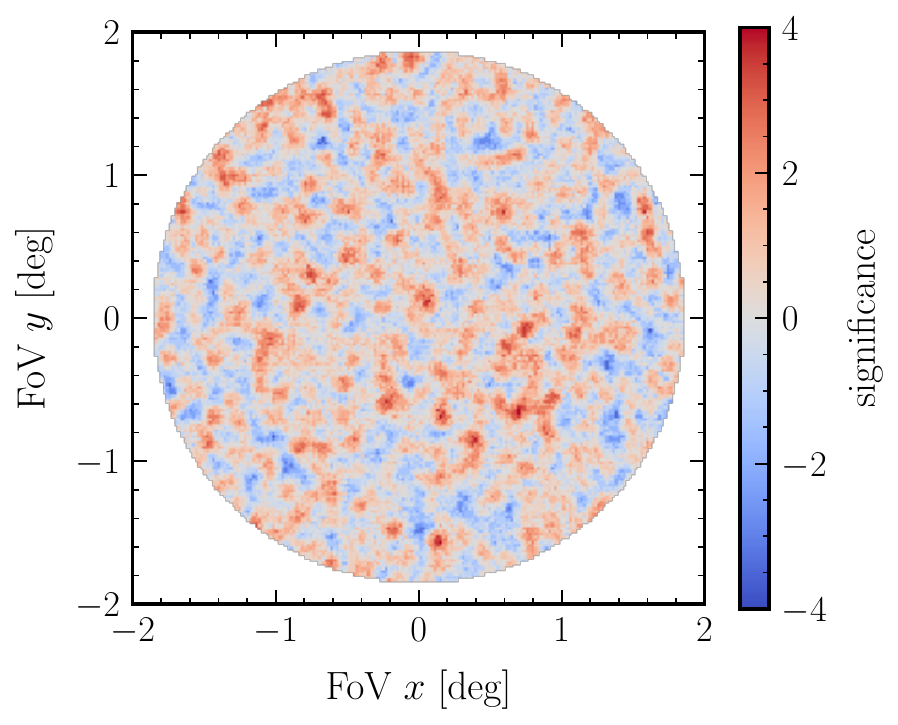}
        \caption{Significance map}
    \end{subfigure}
    \hspace{0.05\textwidth}
    \begin{subfigure}{0.35\textwidth}
        \includegraphics[width=\textwidth]{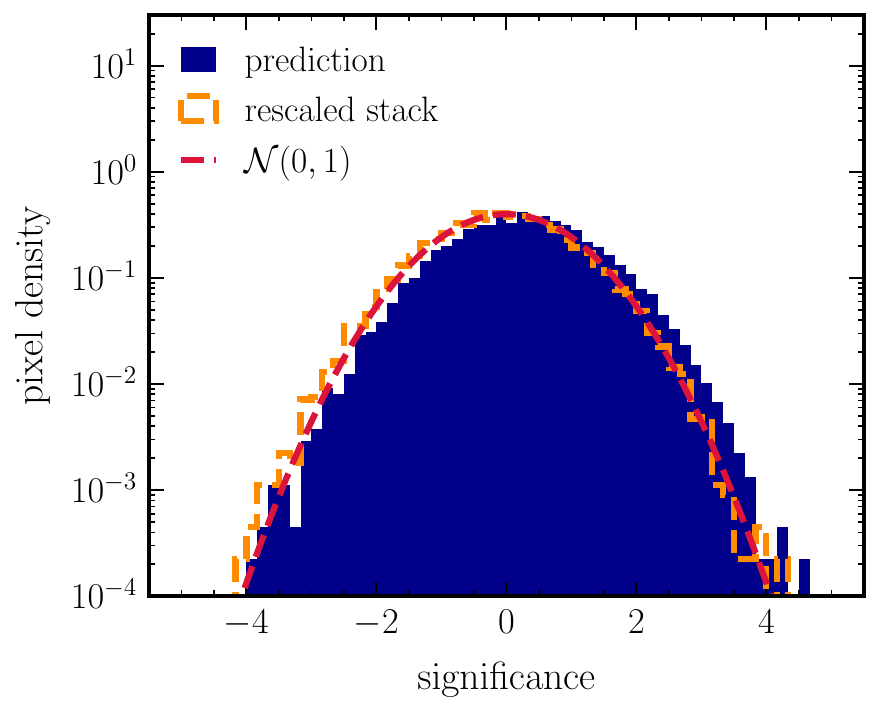}
        \caption{Pixel distributions}
    \end{subfigure}
    \caption{Background of the four Crab runs predicted by the benchmark generative background model, stacked in the FoV frame between $\max(0.5\,\mathrm{TeV},E_\mathrm{safe})$ and 20\,TeV. The source exclusion of $0.4^\circ$ lies at a different position in each run and leaves no pixel excluded in more than half of them. Panel (b) shows the pixel distribution for the prediction ($\mu=+0.24$, $\sigma=1.04$, blue) and after scaling the stacked prediction by a common factor to the observed counts ($\mu=-0.07$, $\sigma=1.02$, orange).}
    \label{fig:crab-bkg}
\end{figure*}

\paragraph{Background of the field.} The four Crab runs (1.8\,h) were taken at zenith angles of 45--48$^\circ$, which is near the upper end of the training pool. The benchmark generative background model underestimates their stacked counts by 12\%, which is 1.6 times the predictive uncertainty of 7\% that the systematic handler assigns to four runs. In fact, the real counts of all four runs lie above the prediction. The significance maps (Fig.~\ref{fig:crab-bkg}) show that this is an offset of the level alone: the distribution is shifted to $\mu=+0.24$, but after scaling the stacked prediction by a common factor to the observed counts, its width of 1.02 is close to the Poisson expectation. The offset is not explained by any known peculiarity of the specific Crab FoV, e.g., by the bright star $\zeta$~Tauri in the field\footnote{$\zeta$~Tauri ($V=3.0$) disables camera pixels during observations of the Crab Nebula \citep{HESS:2006fka}. In the four runs, it lies $0.8$--$2.4^\circ$ from the pointing direction, and the excess of the data is not concentrated along its track through the FoV.}, or by the configuration of the model (Appendix~\ref{app:crab}). With only four runs and without a record of the atmospheric transparency or the camera state, the absolute level of this field cannot be predicted more precisely than the handler states. We therefore infer the background normalization $\eta$ jointly with the source (Sec.~\ref{sec:sbi-analysis}).

\paragraph{Spectrum.} We describe the Crab Nebula as a point source with a power-law spectrum of normalization $N_0$ at 1.45\,TeV and index $\Gamma$, and infer $(N_0,\Gamma,\eta)$ from $10^4$ simulations between 0.871\,TeV, the highest $E_\mathrm{safe}$ of the runs, and 20\,TeV. We note that the Crab is better described by a curved spectrum. However, given that we have only four observation runs at our disposal and we want to compare to previous literature results that also used a power law, we do not add more complexity/realism.

\begin{figure}[t]
    \centering
    \includegraphics[width=\columnwidth]{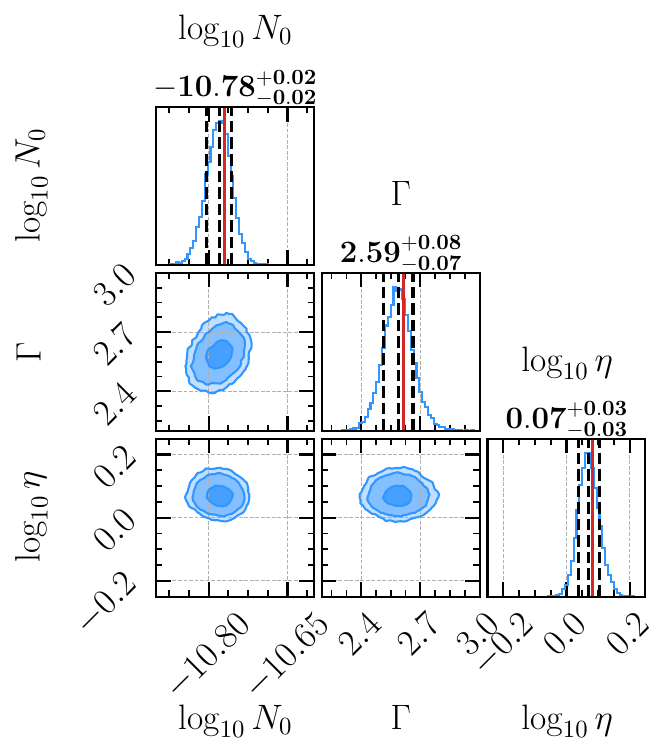}
    \caption{Posterior of the Crab Nebula for a power-law spectrum and an inferred background normalization $\eta$; $N_0$ is given in units of $10^{-12}\,\mathrm{cm^{-2}\,s^{-1}\,TeV^{-1}}$ at 1.45\,TeV. Contours enclose 38\%, 86\% and 95\% of the probability, titles give medians and central 68\% intervals, and red lines mark the classical template fit.}
    \label{fig:crab-post}
\end{figure}

The priors are uniform in $\log_{10}[N_0/(\mathrm{cm^{-2}\,s^{-1}\,TeV^{-1}})]\in[-12.5,-8.5]$, $\Gamma\in[1,4]$ and $\log_{10}\eta\in[-0.25,0.25]$. The posterior (Fig.~\ref{fig:crab-post}) yields $N_0=(16.6\pm0.9)\times10^{-12}\,\mathrm{cm^{-2}\,s^{-1}\,TeV^{-1}}$, $\Gamma=2.59\pm0.08$ and $\eta=1.17\pm0.09$; the inferred normalization matches the offset of the background in this band. In 1000 calibration simulations, the joint coverage lies within $0.64\sigma$ of the nominal levels (Fig.~\ref{fig:crab-calib}), and the ranks of $\Gamma$ and $\eta$ are uniform. The ranks of $N_0$ reveal a small bias of the posterior towards high values, which arises from spectral indices far from that of the Crab Nebula and is absent for $2\leq\Gamma<3$ (Appendix~\ref{app:sbi-val}). The observed counts lie at the center of the posterior predictive distribution ($p=0.86$, relative width 0.10). The classical template fit with a free background normalization gives $N_0=17.0\pm0.8$ and $\Gamma=2.62\pm0.07$ in the same band, which is the vertical red line for comparison in Fig.~\ref{fig:crab-post}.

To compare with the published analyses on this Crab dataset, we need to factor in the different choices on the lower energy threshold. The authors of \citet{Mohrmann:2019hfq} utilized different approaches to determine the spectral parameters at a lower energy threshold between 0.75 and 0.825 TeV: They obtained $N_0=(17.0$--$17.3)\times10^{-12}\,\mathrm{cm^{-2}\,s^{-1}\,TeV^{-1}}$ with $\Gamma=2.66$--$2.69$ for three-dimensional approaches and $N_0=(15.9$--$16.3)\times10^{-12}\,\mathrm{cm^{-2}\,s^{-1}\,TeV^{-1}}$ with $\Gamma=2.59$--$2.63$ using spectral fits only. In contrast, \citet{Wach:2024req} re-analyzed the Crab dataset and found, employing a lower energy threshold of 0.5 TeV, a value of $N_0=(16.1\pm0.7)\times10^{-12}\,\mathrm{cm^{-2}\,s^{-1}\,TeV^{-1}}$ and $\Gamma=2.67\pm0.06$ for their best three-dimensional CR background template. In this sense, our values fall within the range of reported best-fit parameters and they are consistent with them within the stated posterior credible bands. What our SBI approach and the traditional likelihood-based analyses in the literature have in common is that they absorb any CR background normalization offset through the free parameter. However, the SBI posterior additionally propagates its uncertainty into $N_0$.

\subsection{MSH 15$-$52 with SBI}
\label{sec:result-msh}

\paragraph{Background of the field.} The 20 runs of MSH~15$-$52 (8.3\,h) were taken at zenith angles of 35.5--40.6$^\circ$ and at muon efficiencies above those of most training runs (see Sec.~\ref{sec:sbi-analysis}). We again use the benchmark generative background model (Sec.~\ref{sec:bkg-val}), which means that none of the MSH observational runs has been used in its training. Applied to the stacked MSH counts between $\max(0.5\,\mathrm{TeV},E_\mathrm{safe})$ and 20\,TeV and outside a source exclusion of $0.4^\circ$, the model overestimates the observed background events by 9.7\% ($\sim3$ times the uncertainty of 3.4\% that the systematic handler assigns to 20 runs). As for the Crab Nebula, the shape of the background is predicted well, apart from a large-scale gradient that we discuss below, but its total level is not, which may be related to the small amount of training data taken under the conditions of the MSH runs. We therefore again infer the background normalization $\eta$ jointly with the source.  

\paragraph{Simulation, training and parametrization.} In contrast to the Crab Nebula, MSH~15$-$52 is an extended source and, hence, we infer its morphology together with its spectrum. We model the source as a uniform elliptical disc with center (RA, Dec), semi-major axis $a$, eccentricity $e$ and position angle $\phi$ east of north and a power-law spectrum of normalization $N_0$ at 1.4\,TeV and spectral index $\Gamma$. The source's signal convolved with the IRFs (exposure, PSF and energy dispersion at the source position) is regenerated in every simulation via \texttt{gammapy}. Together with $\eta$, eight parameters are inferred, for which we train on $2\times10^4$ simulations between 0.56 and 20\,TeV. The summary network is larger than for the Crab Nebula: we use three trainable convolutional layers with 96 channels, average instead of maximum pooling to a $6\times6$ grid, which preserves the spatial information on elongation and orientation, and a 128-dimensional summary. In addition, the network operates on the native $0.04^\circ$ pixels: unlike for the Crab Nebula, the counts are not summed in blocks of $3\times3$ pixels before entering the network, and the fixed PSF filter uses kernels at the same resolution. We train for a maximum of 200 epochs and select the epoch whose marginal ranks are most uniform on a separate set of 2000 simulations, evaluated after every epoch (epoch 129). Instead of $(e,\phi)$, the flow infers the ellipticity components $e_1=e\cos2\phi$ and $e_2=e\sin2\phi$, as is customary for galaxy shapes in weak lensing \citep{Bartelmann:2001}. The position angle is defined only modulo $180^\circ$, so a posterior close to $0^\circ$ or $180^\circ$ splits into two pieces on the interval $[0^\circ,180^\circ)$, and for a nearly round source $\phi$ is undefined. Even worse, at $e=0$, the coordinates $(e,\phi)$ have a singularity at which any orientation describes the same disc. In the $(e_1,e_2)$-parametrization, the periodicity is explicit, $e=0$ is a regular point and the shape posterior is a single smooth region for any orientation, which a normalizing flow represents far more easily. We choose the prior uniform on the disc $e_1^2+e_2^2\le0.95^2$, i.e.\ isotropic in orientation and $p(e)\propto e$. A prior uniform in $e$ would instead concentrate its density in the shape plane at round shapes, $\propto1/e$, at which the orientation carries no information. The remaining priors are uniform (or uniform in log space) in RA $\in[228.35^\circ,228.75^\circ]$, Dec $\in[-59.35^\circ,-58.99^\circ]$, $a\in[0.08^\circ,0.35^\circ]$, $\Gamma\in[1.5,3.5]$, $\log_{10}[N_0/(\mathrm{cm^{-2}\,s^{-1}\,TeV^{-1}})]\in[-13,-10]$ and $\log_{10}\eta\in[-0.1,0.1]$.

\begin{figure*}[t]
    \centering
    \includegraphics[width=0.8\textwidth]{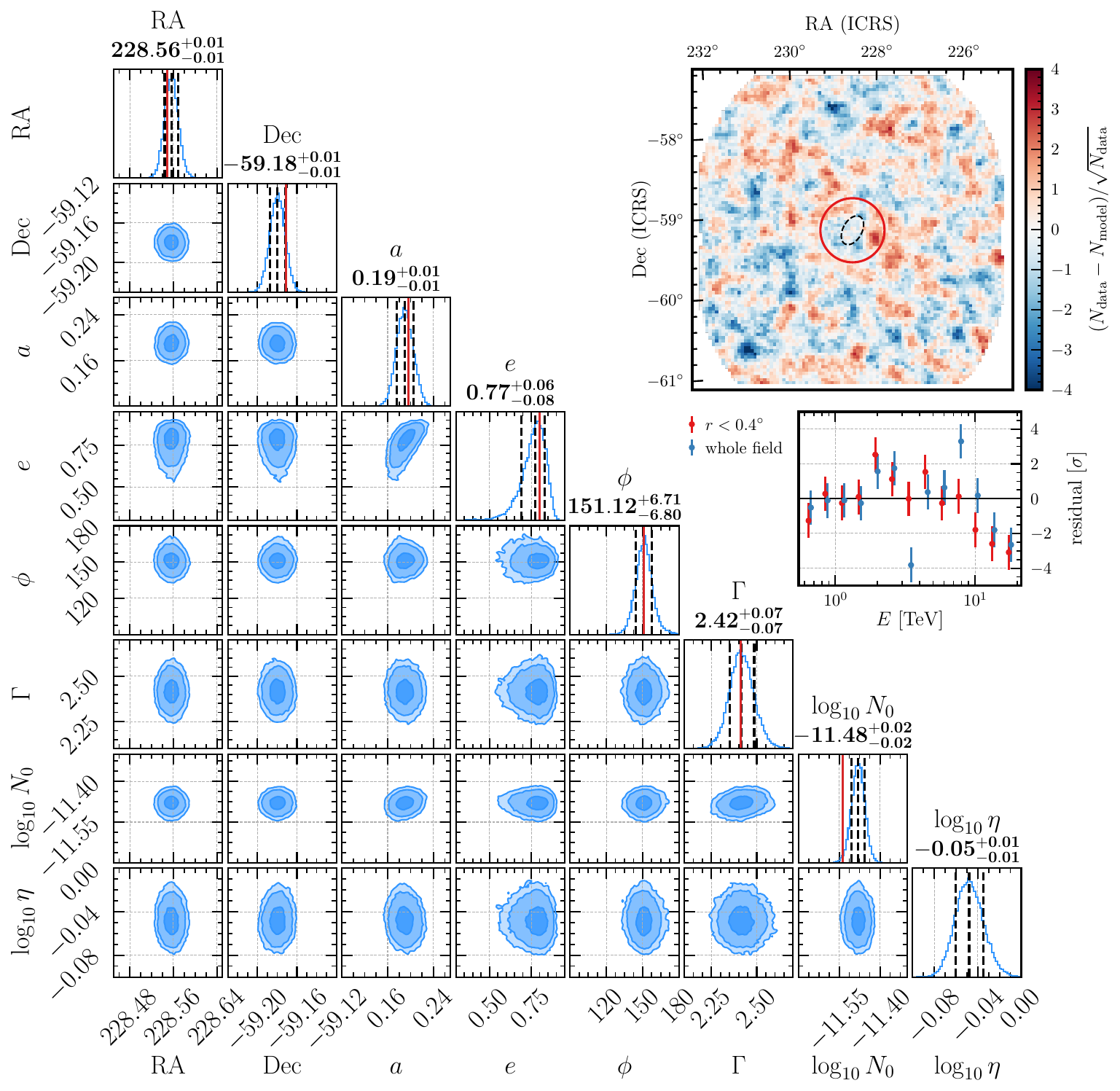}
    \caption{Posterior of MSH~15$-$52 with the benchmark generative background model, for an elliptical uniform disc with a power-law spectrum and an inferred background normalization $\eta$, shown over a reduced prior range for every parameter. The inferred ellipticity components (see text for their definitions) are transformed to the eccentricity $e$ and the position angle $\phi$. RA, Dec and $a$ are in degrees, $N_0$ in $\mathrm{cm^{-2}\,s^{-1}\,TeV^{-1}}$ at 1.4\,TeV. Contours enclose 38\%, 86\% and 95\% of the probability, titles give medians and central 68\% intervals. The vertical red lines in the diagonal plots mark the values of \citet{Mohrmann:2019hfq} (where applicable). Upper right: energy-integrated residuals $(N_{\rm data} - N_{\rm model}) / \sqrt{N_{\rm data}}$, smoothed with a $0.1^{\circ}$ top-hat kernel, in ICRS coordinates. The source is modeled with the MAP parameters; for the CR background, $\eta$ and the run-wise deformations $\bm z_k$ are fitted with the source held fixed (see text). The dashed ellipse marks the MAP shape of MSH~15$-$52. Below: residuals per energy bin (same definition) for the whole field (blue) and within $0.4^{\circ}$ of the MAP center of MSH~15$-$52 (red; red circle in the map).}
    \label{fig:msh-post}
\end{figure*}

\paragraph{Posterior results and calibration checks.} The posterior (Fig.~\ref{fig:msh-post}) locates the source at RA $=228.557^\circ\pm0.013^\circ$ and Dec $=-59.180^\circ\pm0.008^\circ$, with $a=(0.190^{+0.015}_{-0.014})^\circ$, $e=0.77^{+0.06}_{-0.08}$, $\phi=151^\circ\pm7^\circ$, $\Gamma=2.42\pm0.07$ and $N_0=(3.31^{+0.19}_{-0.18})\times10^{-12}\,\mathrm{cm^{-2}\,s^{-1}\,TeV^{-1}}$. The background normalization, $\eta=0.895^{+0.028}_{-0.025}$, is consistent with the 9.7\% overestimate of the background. All parameters are constrained well inside their priors: at most a fraction of $5\times10^{-5}$ of the draws falls into the outer 2\% of any prior range. The contraction of the central 68\% interval relative to the prior is 0.81--0.98 for the position, the size, the spectrum and $\eta$, and 0.72--0.79 for $e_1$ and $e_2$, which the data constrain least. On 1000 calibration simulations (Appendix~\ref{app:sbi-val}), the ranks of RA, Dec, $e_1$ and $\log_{10}N_0$ are uniform ($p\ge0.35$), those of $a$, $e_2$ and $\Gamma$ marginally not ($p=0.011$--0.046), and those of $\log_{10}\eta$ clearly not ($p=6\times10^{-4}$). The marginals are up to 21\% too narrow, and the joint posterior is slightly overconfident: its highest-posterior-density regions (see App.~\ref{app:sbi-val}) cover on average $0.24\sigma$ less than their nominal levels. The observed total of 44\,252 counts lies well within the posterior predictive distribution ($p=0.61$, relative width 0.043). Appendix~\ref{app:msh-val} shows the coverage and the ranks per parameter and discusses how they depend on the selected epoch.

\paragraph{Residuals.} Besides the posterior, Fig.~\ref{fig:msh-post} shows the residuals $(N_{\rm data}-N_{\rm model})/\sqrt{N_{\rm data}}$ of the counts, energy-integrated per pixel (smoothed with a $0.1^\circ$ top-hat) and per energy bin. The source model takes the maximum a posteriori (MAP) parameters, which we obtain by maximizing the NPE density, starting from the best of the posterior draws. For the background, however, a single set of parameters does not represent what the inference has seen: the posterior marginalizes over the run-wise deformations $\bm z_k$ of the systematic handler (Sec.~\ref{sec:sys-handler}), whereas the mean prediction sets all $\bm z_k=0$. With $\bm z_k=0$, the model falls short of the data by 3.0\% and the residuals show a large-scale gradient across the field. We therefore keep the source at the MAP and fit $\eta$ together with the $20\times5$ components of the $\bm z_k$, maximizing the Poisson likelihood times the Gaussian prior of the $\bm z_k$, i.e.\ the conditional MAP of the background nuisance parameters. This removes the gradient. The smoothed spatial residual map has a mean of $-0.08$ and a standard deviation of 1.07, with extremes of $+3.3$ and $-5.4$; the most negative values lie at the edge of the FoV, where few counts bias the statistic towards low values. Per energy bin, the residuals of the whole field stay within $\pm2$ in 10 of 13 bins, but reach $-3.8$ at 3.4\,TeV and $+3.3$ at 7.6\,TeV.

\paragraph{Spectral curvature and emission beyond the disc.} Within $0.4^\circ$ of MSH~15$-$52, the spectral residuals lie between $-1.3$ and $+2.5$ below 10\,TeV and fall to $-1.8$, $-2.6$ and $-3.1$ in the three bins between 10 and 20\,TeV. As a rough estimate utilizing the background residuals in a ring at 0.6--$1.2^\circ$ from the source, we find that about a third of this deficit is due to the background and two thirds due to the source model, whose power law overpredicts the highest energies by about 50\%. On the same public data, \citet{Knodlseder:2019ctools} prefer an exponentially cut-off power law ($\Gamma=1.84\pm0.13$, $E_\mathrm{c}=5.9\pm1.7$\,TeV) over a power law at $4.3\sigma$, which would explain the mismatch we observe by the need for a curved spectrum. Inside the $0.4^\circ$ region, the map shows a single excess of $3.0\sigma$ (pre-trial) at $0.28^\circ$ from the center and at a position angle of $251^\circ$, i.e.\ beyond the edge of the disc along its minor axis with counts spread evenly over 0.6--6\,TeV. Given about 16 independent $0.1^\circ$ cells within $0.4^\circ$, this excess is not significant; it is, however, in line with the larger minor axis that \citet{Knodlseder:2019ctools} obtain for an elliptical Gaussian and attribute to fainter emission on larger scales. A curved spectrum and a morphology beyond a uniform disc are natural extensions of the source model, which we leave to a dedicated study with more exposure on the source.

\paragraph{Galactic emission in the background.} With all deformations at zero, the data exceed the best-fit model by an amount that increases towards the Galactic plane, from $-4\%$ at $b=-3.25^{\circ}$ to $+12\%$  at $b=+0.75^{\circ}$ (outside of $0.4^{\circ}$ from MSH 15--52). The corresponding gradient of $3.4\%$ per degree points within $3^{\circ}\pm14^{\circ}$ of the direction of increasing $b$. When fitted run by run, this gradient is somewhat better described as fixed on the sky than as fixed in the camera frame (about a $2\sigma$ significance), as one would expect for Galactic emission rather than for a mismatch of the acceptance. MSH~15$-$52 lies at $b=-1.2^{\circ}$, and the field extends from $b=-3.9^\circ$ to $+1.5^\circ$. The Galactic plane survey of H.E.S.S.\ \citep[HGPS;][]{HESS:2018HGPS} lists HESS~J1503$-$582 (8\% of the Crab flux above 1\,TeV, $1.7^\circ$ from MSH~15$-$52, $b=+0.3^\circ$) inside the field, as well as HESS~J1457$-$593 and HESS~J1458$-$608 at its edge, and its large-scale component peaks at $b\approx0^\circ$ with a width of about $0.2^\circ$. The observations used to train our benchmark generative background model all point at $|b|\ge4^\circ$ (median $52^\circ$), so the generative model cannot be polluted by Galactic plane emission. \citet{Mohrmann:2019hfq} also report a slight disagreement between their background model and the data in the western part of the field, i.e.\ towards the Galactic plane, which did not affect their source model. Classical analyses absorb such emission implicitly, through the local ring background \citep{HESS:2005MSH} or through per-run gradients of the background template \citep{Knodlseder:2019ctools}. The systematic handler does the same: a deformation common to all runs explains the gradient with tilts of 1.4--$1.6\sigma$ and a normalization of $1.75\sigma$ in units of their prior widths. The generative model can thus accommodate a smooth mismatch of this kind without modeling it explicitly, and since the deformations vary on scales of degrees, they barely affect a source of $0.2^\circ$. The price is that the deformations, meant for the CR background, partly describe astrophysical gamma rays, which biases $\eta$ and the $\bm z_k$. For fields with bright neighbors or structured diffuse emission, known sources and a diffuse component should therefore enter the simulator as additional gamma-ray components or be masked. 

\paragraph{Comparison with BAccMod.} As an independent cross-check of our SBI results, we derive a BAccMod background model that does not rely on any MSH observational run (consistent with the benchmark generative background model). In practice, we build this BAccMod background from the same 55 pool runs as the benchmark generative background model in the setup selected in Sec.~\ref{sec:result-bkg}: for each of the 20 individual MSH runs, a 2D radial model (background rate as a function of offset from the camera center and energy) is constructed from those pool runs taken within $\pm5^\circ$ of that run's zenith angle (amounting to about 8--11 pool runs per MSH run). The resulting background model underestimates the stacked MSH counts by 17\% ($\mu=+1.03$, $\sigma=1.13$), i.e.~it misses the level of this field by more than the generative model but in the opposite direction. With this background, scaled by a free normalization per run, we fit the same disc and power-law model to the per-run counts cubes in the same energy band by maximizing the Poisson likelihood with \texttt{gammapy} (see Table~\ref{tab:msh}); for reference, the table includes the same fit with the mean prediction of the benchmark generative background model ($z_k = 0$) as background template scaled by a free normalization per run. Both classical fits agree with each other as well as with \citet{Mohrmann:2019hfq} in morphology and spectrum. The free normalizations absorb the opposite offsets of the two backgrounds (on average 0.91 and 1.17). The SBI posterior agrees with both classical fits within about one standard deviation in all parameters (the largest difference, in Dec, amounts to $1.1\sigma$ of the combined uncertainties of the SBI and the classical fit with the same background). Its uncertainties are comparable or smaller for the position, $a$, $e$ and the spectrum, but about three times larger for the position angle ($\pm7^\circ$ instead of $\pm2^\circ$). Part of the spatial information is lost in the summary network, whose convolutional layers end in an average pooling to a $6\times6$ grid, whereas the likelihood operates on the full cube. A summary network that keeps more spatial information, such as a transformer, may close this gap. However, we stress that while the classical fits profile a free normalization per run and treat the background shape as exact, the SBI posterior jointly marginalizes over all five calibrated run-to-run systematic modes (rate, radial acceptance, spatial tilts, and spectral tilt), directly propagating instrumental background uncertainty into the inferred source parameters.

We also analyzed the field with backgrounds derived from the MSH runs themselves (with respect to the generative model and with BAccMod). These in-field analyses and their results are reported in Appendix~\ref{app:msh-infield}.

\begin{table}[t]
\centering
\small
\setlength{\tabcolsep}{2.5pt}
\begin{tabular}{lcccc}
\toprule
 & SBI & \multicolumn{2}{c}{classical fit} & M19 \\
 & benchmark & benchmark & BAccMod & \\
\midrule
RA [deg] & $228.557(13)$ & $228.553(25)$ & $228.553(36)$ & $228.549(11)$ \\
Dec [deg] & $-59.180(8)$ & $-59.168(8)$ & $-59.168(14)$ & $-59.171(7)$ \\
$a$ [deg] & $0.190(15)$ & $0.199(17)$ & $0.199(26)$ & $0.196(7)$ \\
$e$ & $0.77^{+0.06}_{-0.08}$ & $0.79(8)$ & $0.79(12)$ & $0.80(4)$ \\
$\phi$ [deg] & $151(7)$ & $154(2)$ & $154(3)$ & $151(6)$ \\
$\Gamma$ & $2.42(7)$ & $2.40(6)$ & $2.35(6)$ & $2.41(5)$ \\
$N_0$ [$10^{-12}$] & $3.31(19)$ & $3.17(16)$ & $3.17(15)$ & $2.91(14)$ \\
$\eta$ & $0.90(3)$ & 0.91 & 1.17 & -- \\
\bottomrule
\end{tabular}
\caption{MSH~15$-$52 for an elliptical uniform disc and a power-law spectrum ($N_0$ in $\mathrm{cm^{-2}\,s^{-1}\,TeV^{-1}}$ at 1.4\,TeV) between 0.56 and 20\,TeV: SBI posterior medians and central 68\% credible intervals with the benchmark generative background model, classical 3D likelihood fits with the same background and with the BAccMod model built from the pool runs (free normalization per run), and \citet[M19]{Mohrmann:2019hfq}. Digits in parentheses give the uncertainty in the last digits. The likelihood of the BAccMod fit is shallow in the morphology so that its quoted uncertainties are only approximate.}
\label{tab:msh}
\end{table}

\section{Conclusion and outlook}
\label{sec:conclusion}

In this work, we have introduced for the first time a unified, data-driven background and inference framework that brings simulation-based inference to real IACT observations. By combining a continuous conditional normalizing flow for the residual CR event density with an out-of-fold calibrated, run-level systematic handler inside a forward simulator, our approach moves beyond binned background templates that lack intrinsic uncertainty models and bridges the simulator gap that has previously confined IACT SBI studies to synthetic data.

We trained this model on the H.E.S.S.~DL3 public test data release~1, using its observations of fields without known gamma-ray emission and, after masking the blazar, runs on PKS~2155$-$304 at high Galactic latitude, which together amount to 55 runs and about 24\,h of livetime. Once trained, the model predicts the expected background of any run whose conditions are covered by the training data, without requiring an emission-free region in its FoV. Since the recorded conditions cannot capture every difference between runs, we complemented the mean prediction with a systematic handler: five smooth deformations of the background intensity, whose run-to-run covariance we calibrated on runs that the respective models had not seen. Of these, the overall level of the background dominates with a scatter of about 15\% between runs, whereas the spatial and spectral shape is reproduced to a few per cent. With Poisson sampling on top, the model becomes a generator of background event lists, which we used as the background component of the forward simulator in SBI analyses of the Crab Nebula and of the extended source MSH~15$-$52, inferring the spectrum and, for MSH~15$-$52, the morphology jointly with the background normalization. Finally, we compared the background predictions and the source parameters with BAccMod models (built from the same runs that enabled the training of the generative background model), with classical likelihood fits, and with published analyses of the same data.

Several aspects of this approach work as intended. First, the conditional flow transfers the spatial and spectral shape of the background to runs it has not seen during training: the Li\&Ma significance distributions obtained from held-out PKS observation runs are as narrow as Poisson fluctuations allow. This statement also applies to the fields of the Crab Nebula and of MSH~15$-$52, once the overall background level is accounted for. In this respect, the model performs at least as well as BAccMod built from the same runs. Second, where the training data cover the field and campaign of a target, the conditioning also predicts the absolute background level: for the held-out runs of PKS~2155$-$304, the stacked prediction matches the observed counts to within its stated uncertainty, whereas the BAccMod model misses them by 12\% unless it is rescaled within the FoV. Third, the systematic handler provides the model with a calibrated statement about its own accuracy, which an SBI analysis can propagate into the posteriors by drawing new run deviations in every simulation. Its smooth deformations absorb large-scale mismatches, such as the gradient of Galactic emission in the field of MSH~15$-$52. Fourth, the complete analysis framework from the background model to calibrated neural posteriors works on real data. For both targets, the SBI posteriors agree with classical likelihood fits and with published analyses of the same runs, including the morphology of MSH~15$-$52. Since the SBI analyses marginalize over all adopted run-to-run systematic uncertainties as well as over the background normalization (where the classical fits profile one normalization per run and neglect the other modes), the SBI uncertainties of the spectral parameters exceed the classical ones by 15--20\%. In contrast, the position and the size of the source, which the smooth deformations cannot mimic, are constrained equally well. The larger uncertainties are the expected consequence of a residual CR background whose accuracy is part of the model and directly enters the analysis instead of being absorbed into free normalizations, as in traditional likelihood fits.

The main limitation of the present model is the absolute level of the background in fields that are absent from the training data, while the shape of the background is reproduced. For the Crab Nebula and for MSH~15$-$52, the prediction misses the observed counts by $-12\%$ and $+10\%$, respectively. This is more than the systematic handler allows for. We therefore had to infer the background normalization jointly with the source, so that the level is effectively taken from the target field itself, as in classical analyses, and emission that is uniform across the FoV remains degenerate with it. We attribute this limitation to the small training set rather than to the method. The public release contains 55 usable runs from a few distinct fields; the target runs lie at the edges of the sampled conditions, at large zenith angles for the Crab Nebula and at high optical efficiencies for MSH~15$-$52. Additionally, the per-run metadata lack quantities that are known to influence the background level, notably the atmospheric transparency, the night-sky background and the state of the cameras. 

The same scarcity of data limits the systematic handler: with few independent fields, we can only describe run-to-run deviations that are independent and centered on zero, whereas deviations shared by the runs of one field or a residual bias of the model as a function of the conditions would require many more fields to be calibrated. A weak radial residual common to all runs (that can also be identified in a BAccMod model built from the same runs) points to a deviation of the mean intensity that the handler cannot describe. All these issues are expected to diminish when the model is trained on the archival data of an IACT collaboration, which contain substantially larger combined livetimes. Moreover, relevant additional conditions and monitoring information, such as atmospheric properties per observation run, trigger rates or the fraction of active pixels, are certainly adding predictive power to the generative model. Finally, the model describes only what is fixed in the camera frame. Bright stars and astrophysical emission that are linked to positions in the sky must be masked or modeled separately. Otherwise, the smooth deformations of the systematic handler absorb them (as the Galactic emission in the field of MSH~15$-$52 shows) and bias the inferred background parameters.

On the inference side, the neural posteriors do not yet meet all of our calibration criteria. The posterior of the Crab Nebula is calibrated only locally around the observed spectral index, and that of MSH~15$-$52 is slightly overconfident, with the background normalization being the least calibrated parameter. We also have seen that joint coverage and marginal ranks reach their optima at different stages of the training. Established strategies to remedy these small deficiencies exist: larger training sets, ensembles of flows (that also capture the epistemic uncertainty intrinsic to stochastically initialized neural networks), training objectives that penalize overconfidence or an epoch selection that requires both calibration metrics. The convolutional summary network, in turn, discards part of the spatial information, which shows up as a three times larger uncertainty of the position angle of MSH~15$-$52 than in the classical fit. Architectures that retain more spatial detail, such as transformers, may close this gap. Lastly, the source models we used are deliberately simple and essentially adopted to enable a comparison with literature results on the same public dataset. The residuals of MSH~15$-$52, e.g.,  point to a curved spectrum and potentially to emission beyond the uniform disc. In science-driven analyses, one would rather employ test statistics that are sensitive to spectral and morphological residuals to render such mis-specifications visible within the SBI framework itself.

Once trained on larger archives with richer atmospheric and instrumental monitoring (removing the need to refit the background normalization in-field), a systematics-informed generative model will be most valuable where background uncertainties limit sensitivity. This spans deep exposures of faint targets and searches for new physics, such as dark matter annihilation in dwarf spheroidal galaxies or the Galactic Center and spectral signatures of axion-like particles \citep{Bhattacharjee:2025ofd}, as well as FoV-filling sources like large supernova remnants, pulsar halos \citep{HAWC:2017kbo, Eckner:2022gil}, and crowded regions of the Galactic plane \citep{Mestre:2022kip}. In such crowded fields, any in-field background estimator (including a field-trained generative model; Appendix~\ref{app:msh-infield}) absorbs unmodeled diffuse or extended emission, whereas high-latitude models avoid this self-subtraction, as illustrated by the Galactic latitude gradient exposed in MSH~15$-$52. Beyond high-latitude training alone, our systematic handler separates astrophysical signal from instrumental mismodeling: because its calibrated deformations act strictly in the camera frame and vary between runs, persistent sky-fixed emission resists absorption as exposure accumulates and can be modeled explicitly within the SBI forward simulator.

Because this approach requires only standard DL3 event lists and per-run metadata, it transfers directly from H.E.S.S.~to MAGIC, VERITAS, and LST-1 of CTAO \citep{CTA-LSTProject:2023haa}, as well as to joint multi-instrument analyses in which each array enters the forward simulator with its own background model. Existing collaborations already hold decades of archival observations paired with detailed environmental monitoring, such as the LiDAR atmospheric profiling at the MAGIC site; unlike binned templates, a conditional normalizing flow shares statistical strength continuously across runs and natively conditions on any such monitoring quantity. Crucially, this framework will scale with CTAO from day one: during the early science phase of the LST subarray at CTAO-North, the systematic handler quantifies background uncertainty while archival statistics are modest, and periodic retraining tightens both predictions and uncertainties as the archive grows. By turning continuous monitoring and growing observation archives into a unified, systematics-aware forward model, generative CR backgrounds offer a scalable path to controlling background uncertainties, unlocking the full legacy of current IACTs and meeting the precision-science demands of CTAO.

\begin{acknowledgements}
This work made use of data from the H.E.S.S. DL3 public test data release 1 (HESS DL3 DR1, \cite{HESS:2018zix}).\\
The data production, processing and analysis tools for this paper have been developed, implemented and operated in collaboration with the Port d’Informació Científica (PIC) data center. PIC is maintained through a collaboration agreement between the Institut de Física d’Altes Energies (IFAE) and the Centro de Investigaciones Energéticas, Medioambientales y Tecnológicas (CIEMAT).\\
Parts of the analysis code and figures in this work were developed with the assistance of \textit{Claude Code (Opus 5.5)}. All AI-produced output was checked, reviewed and edited by the authors, who take full responsibility for the content.

The work of CE, FC and JMC is supported by the European Union through the grant ``UNDARK'' of the Widening participation and spreading excellence programme (project number 101159929).

The work of FC, and SC is supported by the Agence Nationale de la Recherche through the project ANR-25-CE31-6485.
\end{acknowledgements}

\bibliographystyle{aa}
\bibliography{refsdraft}

@article{HESS:2018zix,
    author = "Abdalla, H. and others",
    collaboration = "H.E.S.S.",
    title = "{H.E.S.S. first public test data release}",
    eprint = "1810.04516",
    archivePrefix = "arXiv",
    primaryClass = "astro-ph.HE",
    doi = "10.5281/zenodo.1421098",
    month = "10",
    year = "2018"
}

@inproceedings{Rezende:2015ocs,
    author = "Rezende, Danilo Jimenez and Mohamed, Shakir",
    title = "{Variational Inference with Normalizing Flows}",
    eprint = "1505.05770",
    archivePrefix = "arXiv",
    primaryClass = "stat.ML",
    month = "5",
    year = "2015",
    booktitle = "{Proceedings of the 32nd International Conference on Machine Learning, PMLR 37}"
}

@article{Kobyzev:2019ydm,
    author = "Kobyzev, Ivan and Prince, Simon J. D. and Brubaker, Marcus A.",
    title = "{Normalizing Flows: An Introduction and Review of Current Methods}",
    eprint = "1908.09257",
    archivePrefix = "arXiv",
    primaryClass = "stat.ML",
    doi = "10.1109/tpami.2020.2992934",
    journal = "IEEE Trans. Pattern Anal. Machine Intell.",
    volume = "43",
    number = "11",
    pages = "3964--3979",
    year = "2021"
}

@article{Papamakarios:2019fms,
    author = "Papamakarios, George and Nalisnick, Eric and Rezende, Danilo Jimenez and Mohamed, Shakir and Lakshminarayanan, Balaji",
    title = "{Normalizing Flows for Probabilistic Modeling and Inference}",
    eprint = "1912.02762",
    archivePrefix = "arXiv",
    primaryClass = "stat.ML",
    doi = "10.5555/3546258.3546315",
    journal = "J. Machine Learning Res.",
    volume = "22",
    number = "1",
    pages = "2617--2680",
    year = "2021"
}

@article{Winkler:2019lio,
    author = "Winkler, Christina and Worrall, Daniel and Hoogeboom, Emiel and Welling, Max",
    title = "{Learning Likelihoods with Conditional Normalizing Flows}",
    eprint = "1912.00042",
    archivePrefix = "arXiv",
    primaryClass = "cs.LG",
    month = "11",
    year = "2019"
}

@inproceedings{Papamakarios:2017tec,
    author = "Papamakarios, George and Pavlakou, Theo and Murray, Iain",
    title = "{Masked Autoregressive Flow for Density Estimation}",
    eprint = "1705.07057",
    archivePrefix = "arXiv",
    primaryClass = "stat.ML",
    month = "5",
    year = "2017",
    booktitle = "{Advances in Neural Information Processing Systems 30}"
}

@inproceedings{Germain:2015yft,
    author = "Germain, Mathieu and Gregor, Karol and Murray, Iain and Larochelle, Hugo",
    title = "{MADE: Masked Autoencoder for Distribution Estimation}",
    eprint = "1502.03509",
    archivePrefix = "arXiv",
    primaryClass = "cs.LG",
    month = "2",
    year = "2015",
    booktitle = "{Proceedings of the 32nd International Conference on Machine Learning, PMLR 37}"
}

@inproceedings{Durkan:2019nsq,
    author = "Durkan, Conor and Bekasov, Artur and Murray, Iain and Papamakarios, George",
    title = "{Neural Spline Flows}",
    eprint = "1906.04032",
    archivePrefix = "arXiv",
    primaryClass = "stat.ML",
    month = "6",
    year = "2019",
    booktitle = "{Advances in Neural Information Processing Systems 32}"
}

@article{Mohrmann:2019hfq,
    author = "Mohrmann, Lars and Specovius, Andreas and Tiziani, Domenico and Funk, Stefan and Malyshev, Dmitry and Nakashima, Kaori and van Eldik, Christopher",
    title = "{Validation of open-source science tools and background model construction in $\gamma$-ray astronomy}",
    eprint = "1910.08088",
    archivePrefix = "arXiv",
    primaryClass = "astro-ph.IM",
    doi = "10.1051/0004-6361/201936452",
    journal = "Astron. Astrophys.",
    volume = "632",
    pages = "A72",
    year = "2019"
}

@article{Cranmer:2019eaq,
    author = "Cranmer, Kyle and Brehmer, Johann and Louppe, Gilles",
    title = "{The frontier of simulation-based inference}",
    eprint = "1911.01429",
    archivePrefix = "arXiv",
    primaryClass = "stat.ML",
    doi = "10.1073/pnas.1912789117",
    journal = "Proc. Nat. Acad. Sci.",
    volume = "117",
    number = "48",
    pages = "30055--30062",
    year = "2020"
}

@article{Nachman:2020lpy,
    author = "Nachman, Benjamin and Shih, David",
    title = "{Anomaly Detection with Density Estimation}",
    eprint = "2001.04990",
    archivePrefix = "arXiv",
    primaryClass = "hep-ph",
    doi = "10.1103/PhysRevD.101.075042",
    journal = "Phys. Rev. D",
    volume = "101",
    pages = "075042",
    year = "2020"
}

@article{Hallin:2021wme,
    author = "Hallin, Anna and Isaacson, Joshua and Kasieczka, Gregor and Krause, Claudius and Nachman, Benjamin and Quadfasel, Tobias and Schlaffer, Matthias and Shih, David and Sommerhalder, Manuel",
    title = "{Classifying anomalies through outer density estimation}",
    eprint = "2109.00546",
    archivePrefix = "arXiv",
    primaryClass = "hep-ph",
    reportNumber = "EFI-20-5, FERMILAB-PUB-21-389-T",
    doi = "10.1103/PhysRevD.106.055006",
    journal = "Phys. Rev. D",
    volume = "106",
    number = "5",
    pages = "055006",
    year = "2022"
}

@article{Berge:2006ae,
    author = "Berge, David and Funk, S. and Hinton, J.",
    title = "{Background Modelling in Very-High-Energy gamma-ray Astronomy}",
    eprint = "astro-ph/0610959",
    archivePrefix = "arXiv",
    reportNumber = "SLAC-PUB-12185",
    doi = "10.1051/0004-6361:20066674",
    journal = "Astron. Astrophys.",
    volume = "466",
    pages = "1219--1229",
    year = "2007"
}

@inproceedings{Chalme-Calvet:2014qxa,
    author = "Chalme-Calvet, R. and de Naurois, M. and Tavernet, J. -P.",
    collaboration = "H.E.S.S.",
    title = "{Muon efficiency of the H.E.S.S. telescope}",
    booktitle = "{International Workshop on Atmospheric Monitoring for High-Energy Astroparticle Detectors}",
    eprint = "1403.4550",
    archivePrefix = "arXiv",
    primaryClass = "astro-ph.IM",
    month = "3",
    year = "2014"
}

@article{Wach:2024req,
    author = "Wach, Tina and Mitchell, Alison and Mohrmann, Lars",
    title = "{A background-estimation technique for the detection of extended gamma-ray structures with IACTs}",
    eprint = "2409.02527",
    archivePrefix = "arXiv",
    primaryClass = "astro-ph.IM",
    doi = "10.1051/0004-6361/202451020",
    journal = "Astron. Astrophys.",
    volume = "690",
    pages = "A250",
    year = "2024"
}

@article{Barlow:1990vc,
    author = "Barlow, Roger J.",
    title = "{Extended maximum likelihood}",
    reportNumber = "M-C-EXP-90-1",
    doi = "10.1016/0168-9002(90)91334-8",
    journal = "Nucl. Instrum. Meth. A",
    volume = "297",
    pages = "496--506",
    year = "1990"
}

@article{Cash:1979vz,
    author = "Cash, W.",
    title = "{Parameter estimation in astronomy through application of the likelihood ratio}",
    doi = "10.1086/156922",
    journal = "Astrophys. J.",
    volume = "228",
    pages = "939--947",
    year = "1979"
}

@article{Hendrycks:2016qxa,
    author = "Hendrycks, Dan and Gimpel, Kevin",
    title = "{Gaussian Error Linear Units (GELUs)}",
    eprint = "1606.08415",
    archivePrefix = "arXiv",
    primaryClass = "cs.LG",
    month = "6",
    year = "2016"
}

@inproceedings{Paszke:2019xhz,
    author = "Paszke, Adam and others",
    title = "{PyTorch: An Imperative Style, High-Performance Deep Learning Library}",
    eprint = "1912.01703",
    archivePrefix = "arXiv",
    primaryClass = "cs.LG",
    month = "12",
    year = "2019",
    booktitle = "{Advances in Neural Information Processing Systems 32}"
}

@article{Chadwick:1999hx,
    author = "Chadwick, P. M. and Lyons, K. and McComb, T. JL. and Orford, K. J. and Osborne, J. L. and Rayner, S. M. and Roberts, I. D. and Shaw, S. E. and Turver, K. E.",
    title = "{Geomagnetic effects on atmospheric Cherenkov images}",
    eprint = "astro-ph/9904082",
    archivePrefix = "arXiv",
    reportNumber = "DUR-PHYS-VHE-98-09",
    doi = "10.1088/0954-3899/25/6/312",
    journal = "J. Phys. G",
    volume = "25",
    pages = "1223--1233",
    year = "1999"
}

@article{Commichau:2008pe,
    author = "Commichau, S. C. and Biland, A. and Contreras, J. L. and de los Reyes, R. and Moralejo, A. and Sitarek, J. and Sobczynska, D.",
    title = "{Monte Carlo Studies of Geomagnetic Field Effects on the Imaging Air Cherenkov Technique}",
    eprint = "0802.2551",
    archivePrefix = "arXiv",
    primaryClass = "astro-ph",
    doi = "10.1016/j.nima.2008.07.144",
    journal = "Nucl. Instrum. Meth. A",
    volume = "595",
    pages = "572--586",
    year = "2008"
}

@article{Bernlohr:2000wq,
    author = "Bernlohr, K.",
    title = "{Impact of atmospheric parameters on the atmospheric Cherenkov technique}",
    eprint = "astro-ph/9908093",
    archivePrefix = "arXiv",
    doi = "10.1016/S0927-6505(99)00093-6",
    journal = "Astropart. Phys.",
    volume = "12",
    pages = "255--268",
    year = "2000"
}

@article{Hahn:2013pfa,
    author = "Hahn, J. and others",
    title = "{Impact of aerosols and adverse atmospheric conditions on the data quality for spectral analysis of the H.E.S.S. telescopes}",
    eprint = "1310.1639",
    archivePrefix = "arXiv",
    primaryClass = "astro-ph.IM",
    doi = "10.1016/j.astropartphys.2013.10.003",
    journal = "Astropart. Phys.",
    volume = "54",
    pages = "25--32",
    year = "2014"
}

@article{MAGIC:2017zph,
    author = "Ahnen, M. L. and others",
    collaboration = "MAGIC",
    title = "{Performance of the MAGIC telescopes under moonlight}",
    eprint = "1704.00906",
    archivePrefix = "arXiv",
    primaryClass = "astro-ph.IM",
    doi = "10.1016/j.astropartphys.2017.08.001",
    journal = "Astropart. Phys.",
    volume = "94",
    pages = "29--41",
    year = "2017"
}

@article{Archambault:2017jgj,
    author = "Archambault, S. and others",
    title = "{Gamma-ray observations under bright moonlight with VERITAS}",
    eprint = "1703.01307",
    archivePrefix = "arXiv",
    primaryClass = "astro-ph.IM",
    doi = "10.1016/j.astropartphys.2017.03.001",
    journal = "Astropart. Phys.",
    volume = "91",
    pages = "34--43",
    year = "2017"
}

@article{Eckner:2025waa,
    author = "Eckner, Christopher and Anau Montel, Noemi and List, Florian and Calore, Francesca and Weniger, Christoph",
    title = "{Robust neural determination of the source-count distribution of the Fermi-LAT sky at high latitudes}",
    eprint = "2505.02906",
    archivePrefix = "arXiv",
    primaryClass = "astro-ph.HE",
    reportNumber = "LAPTH-016/25",
    doi = "10.1103/p4gf-d2zt",
    journal = "Phys. Rev. D",
    volume = "112",
    number = "10",
    pages = "103022",
    year = "2025"
}

@book{Kingman:1993,
    author = "Kingman, J. F. C.",
    title = "{Poisson Processes}",
    series = "Oxford Studies in Probability",
    volume = "3",
    publisher = "Clarendon Press",
    address = "Oxford",
    year = "1993"
}

@book{Daley:2003,
    author = "Daley, D. J. and Vere-Jones, D.",
    title = "{An Introduction to the Theory of Point Processes. Volume I: Elementary Theory and Methods}",
    edition = "2nd",
    series = "Probability and its Applications",
    publisher = "Springer",
    address = "New York",
    doi = "10.1007/b97277",
    year = "2003"
}

@article{Cox:1955,
    author = "Cox, D. R.",
    title = "{Some Statistical Methods Connected with Series of Events}",
    journal = "J. R. Stat. Soc. B",
    volume = "17",
    number = "2",
    pages = "129--157",
    doi = "10.1111/j.2517-6161.1955.tb00188.x",
    year = "1955"
}

@article{Moller:1998,
    author = "M{\o}ller, J. and Syversveen, A. R. and Waagepetersen, R. P.",
    title = "{Log Gaussian Cox Processes}",
    journal = "Scand. J. Stat.",
    volume = "25",
    number = "3",
    pages = "451--482",
    doi = "10.1111/1467-9469.00115",
    year = "1998"
}

@article{Gourieroux:1984,
    author = "Gourieroux, C. and Monfort, A. and Trognon, A.",
    title = "{Pseudo Maximum Likelihood Methods: Applications to Poisson Models}",
    journal = "Econometrica",
    volume = "52",
    number = "3",
    pages = "701--720",
    doi = "10.2307/1913472",
    year = "1984"
}

@misc{Rozet:2022zuko,
    author = "Rozet, Fran{\c{c}}ois and Divo, Felix and Schnake, Simon",
    title = "{Zuko: Normalizing flows in PyTorch}",
    howpublished = "Zenodo",
    doi = "10.5281/zenodo.7625672",
    year = "2022"
}

@article{HESS:2006fka,
    author = "Aharonian, F. and others",
    collaboration = "H.E.S.S.",
    title = "{Observations of the Crab Nebula with H.E.S.S}",
    eprint = "astro-ph/0607333",
    archivePrefix = "arXiv",
    doi = "10.1051/0004-6361:20065351",
    journal = "Astron. Astrophys.",
    volume = "457",
    pages = "899--915",
    year = "2006"
}

@article{Li:1983fv,
    author = "Li, T. -P. and Ma, Y. -Q.",
    title = "{Analysis methods for results in gamma-ray astronomy}",
    doi = "10.1086/161295",
    journal = "Astrophys. J.",
    volume = "272",
    pages = "317--324",
    year = "1983"
}

@article{Gammapy:2023gvb,
    author = "Donath, Axel and others",
    collaboration = "Gammapy",
    title = "{Gammapy: A Python package for gamma-ray astronomy}",
    eprint = "2308.13584",
    archivePrefix = "arXiv",
    primaryClass = "astro-ph.IM",
    doi = "10.1051/0004-6361/202346488",
    journal = "Astron. Astrophys.",
    volume = "678",
    pages = "A157",
    year = "2023"
}

@article{Talts:2018zdk,
    author = "Talts, Sean and Betancourt, Michael and Simpson, Daniel and Vehtari, Aki and Gelman, Andrew",
    title = "{Validating Bayesian Inference Algorithms with Simulation-Based Calibration}",
    eprint = "1804.06788",
    archivePrefix = "arXiv",
    primaryClass = "stat.ME",
    month = "4",
    year = "2018"
}

@article{Hermans:2021rqv,
    author = "Hermans, Joeri and Delaunoy, Arnaud and Rozet, Fran{\c{c}}ois and Wehenkel, Antoine and Begy, Volodimir and Louppe, Gilles",
    title = "{A Trust Crisis In Simulation-Based Inference? Your Posterior Approximations Can Be Unfaithful}",
    eprint = "2110.06581",
    archivePrefix = "arXiv",
    primaryClass = "stat.ML",
    month = "10",
    year = "2021"
}

@inproceedings{Kingma:2014vow,
    author = "Kingma, Diederik P. and Ba, Jimmy",
    title = "{Adam: A Method for Stochastic Optimization}",
    booktitle = "{International Conference on Learning Representations}",
    eprint = "1412.6980",
    archivePrefix = "arXiv",
    primaryClass = "cs.LG",
    month = "12",
    year = "2014"
}

@article{Papamakarios:2016ctj,
    author = "Papamakarios, George and Murray, Iain",
    title = "{Fast $\epsilon$-free Inference of Simulation Models with Bayesian Conditional Density Estimation}",
    eprint = "1605.06376",
    archivePrefix = "arXiv",
    primaryClass = "stat.ML",
    month = "5",
    year = "2016"
}

@inproceedings{Miller:2021hys,
    author = "Miller, Benjamin Kurt and Cole, Alex and Forr{\'e}, Patrick and Louppe, Gilles and Weniger, Christoph",
    title = "{Truncated Marginal Neural Ratio Estimation}",
    booktitle = "{35th Conference on Neural Information Processing Systems}",
    eprint = "2107.01214",
    archivePrefix = "arXiv",
    primaryClass = "stat.ML",
    doi = "10.5281/zenodo.5043706",
    month = "7",
    year = "2021"
}

@misc{deBonydeLavergne:2026baccmod,
    author = "{de Bony de Lavergne}, Mathieu and Emery, Gabriel and Carrasco, Marie-Sophie and Aguasca-Cabot, Arnau and Fr{\"o}se, Stefan",
    title = "{BAccMod}",
    howpublished = "Zenodo",
    doi = "10.5281/zenodo.17256957",
    note = "Version 0.4.1, \url{https://github.com/mdebony/BAccMod}",
    year = "2026"
}

@article{Cook:2006,
    author = "Cook, Samantha R. and Gelman, Andrew and Rubin, Donald B.",
    title = "{Validation of Software for Bayesian Models Using Posterior Quantiles}",
    journal = "J. Comput. Graph. Stat.",
    volume = "15",
    pages = "675--692",
    doi = "10.1198/106186006X136976",
    year = "2006"
}

@article{Hyndman:1996,
    author = "Hyndman, Rob J.",
    title = "{Computing and Graphing Highest Density Regions}",
    journal = "Am. Stat.",
    volume = "50",
    pages = "120--126",
    doi = "10.1080/00031305.1996.10474359",
    year = "1996"
}

@article{Brown:2001,
    author = "Brown, Lawrence D. and Cai, T. Tony and DasGupta, Anirban",
    title = "{Interval Estimation for a Binomial Proportion}",
    journal = "Stat. Sci.",
    volume = "16",
    pages = "101--133",
    doi = "10.1214/ss/1009213286",
    year = "2001"
}

@article{Rubin:1984,
    author = "Rubin, Donald B.",
    title = "{Bayesianly Justifiable and Relevant Frequency Calculations for the Applied Statistician}",
    journal = "Ann. Stat.",
    volume = "12",
    pages = "1151--1172",
    doi = "10.1214/aos/1176346785",
    year = "1984"
}

@article{Gelman:1996,
    author = "Gelman, Andrew and Meng, Xiao-Li and Stern, Hal",
    title = "{Posterior Predictive Assessment of Model Fitness via Realized Discrepancies}",
    journal = "Stat. Sin.",
    volume = "6",
    pages = "733--807",
    year = "1996"
}

@article{Bartelmann:2001,
    author = "Bartelmann, Matthias and Schneider, Peter",
    title = "{Weak gravitational lensing}",
    journal = "Phys. Rep.",
    volume = "340",
    pages = "291--472",
    doi = "10.1016/S0370-1573(00)00082-X",
    eprint = "astro-ph/9912508",
    archivePrefix = "arXiv",
    year = "2001"
}

@inproceedings{Lakshminarayanan:2017,
    author = "Lakshminarayanan, Balaji and Pritzel, Alexander and Blundell, Charles",
    title = "{Simple and Scalable Predictive Uncertainty Estimation using Deep Ensembles}",
    booktitle = "{Advances in Neural Information Processing Systems 30}",
    eprint = "1612.01474",
    archivePrefix = "arXiv",
    year = "2017"
}

@inproceedings{Delaunoy:2022,
    author = "Delaunoy, Arnaud and Hermans, Joeri and Rozet, Fran{\c{c}}ois and Wehenkel, Antoine and Louppe, Gilles",
    title = "{Towards Reliable Simulation-Based Inference with Balanced Neural Ratio Estimation}",
    booktitle = "{Advances in Neural Information Processing Systems 35}",
    eprint = "2208.13624",
    archivePrefix = "arXiv",
    year = "2022"
}

@article{Delaunoy:2023,
    author = "Delaunoy, Arnaud and Miller, Benjamin Kurt and Forr{\'e}, Patrick and Weniger, Christoph and Louppe, Gilles",
    title = "{Balancing Simulation-based Inference for Conservative Posteriors}",
    eprint = "2304.10978",
    archivePrefix = "arXiv",
    primaryClass = "stat.ML",
    year = "2023"
}

@inproceedings{Falkiewicz:2023,
    author = "Falkiewicz, Maciej and Takeishi, Naoya and Shekhzadeh, Imahn and Wehenkel, Antoine and Delaunoy, Arnaud and Louppe, Gilles and Kalousis, Alexandros",
    title = "{Calibrating Neural Simulation-Based Inference with Differentiable Coverage Probability}",
    booktitle = "{Advances in Neural Information Processing Systems 36}",
    eprint = "2310.13402",
    archivePrefix = "arXiv",
    year = "2023"
}

@INPROCEEDINGS{1985ICRC....3..445H,
       author = {{Hillas}, A.~M.},
        title = "{Cerenkov Light Images of EAS Produced by Primary Gamma Rays and by Nuclei}",
    booktitle = {19th International Cosmic Ray Conference (ICRC19), Volume 3},
         year = 1985,
       editor = {{Jones}, F.~C.},
       series = {International Cosmic Ray Conference},
       volume = {3},
        month = aug,
        pages = {445},
       adsurl = {https://ui.adsabs.harvard.edu/abs/1985ICRC....3..445H}
}

@ARTICLE{1989ApJ...342..379W,
       author = {{Weekes}, T.~C. and {Cawley}, M.~F. and {Fegan}, D.~J. and {Gibbs}, K.~G. and {Hillas}, A.~M. and {Kowk}, P.~W. and {Lamb}, R.~C. and {Lewis}, D.~A. and {Macomb}, D. and {Porter}, N.~A. and {Reynolds}, P.~T. and {Vacanti}, G.},
        title = "{Observation of TeV Gamma Rays from the Crab Nebula Using the Atmospheric Cerenkov Imaging Technique}",
      journal = "{Astrophys.~J.}",
         year = 1989,
        month = jul,
       volume = {342},
        pages = {379},
          doi = {10.1086/167599},
       adsurl = {https://ui.adsabs.harvard.edu/abs/1989ApJ...342..379W}
}

@ARTICLE{1994APh.....2..137F,
       author = {{Fomin}, V.~P. and {Stepanian}, A.~A. and {Lamb}, R.~C. and {Lewis}, D.~A. and {Punch}, M. and {Weekes}, T.~C.},
        title = "{New methods of atmospheric Cherenkov imaging for gamma-ray astronomy. I. The false source method}",
      journal = {Astroparticle Physics},
         year = 1994,
        month = may,
       volume = {2},
       number = {2},
        pages = {137-150},
          doi = {10.1016/0927-6505(94)90036-1},
       adsurl = {https://ui.adsabs.harvard.edu/abs/1994APh.....2..137F}
}

@article{Aharonian:2001mz,
    author = "Aharonian, F.",
    collaboration = "HEGRA",
    title = "{Evidence for TeV gamma-ray emission from Cassiopeia A}",
    eprint = "astro-ph/0102391",
    archivePrefix = "arXiv",
    doi = "10.1051/0004-6361:20010243",
    journal = "Astron. Astrophys.",
    volume = "370",
    number = "1",
    pages = "112--120",
    year = "2001"
}

@article{Demichev:2024mnc,
    author = "Demichev, A. and Kryukov, A.",
    title = "{Using deep learning methods for IACT data analysis in gamma-ray astronomy: A review}",
    doi = "10.1016/j.ascom.2024.100793",
    journal = "Astron. Comput.",
    volume = "46",
    pages = "100793",
    year = "2024"
}

@article{Kryukov:2024hka,
    author = "Kryukov, A. P. and Demichev, A. P. and Ilyin, V. A.",
    title = "{Machine Learning in Gamma Astronomy}",
    eprint = "2501.19064",
    archivePrefix = "arXiv",
    doi = "10.3103/S002713492470214X",
    journal = "Moscow Univ. Phys. Bull.",
    volume = "79",
    number = "Suppl 2",
    pages = "S690--S699",
    year = "2024"
}

@article{Bock:2004td,
    author = "Bock, R. K. and Chilingarian, A. and Gaug, M. and Hakl, F. and Hengstebeck, T. and Jirina, M. and Klaschka, J. and Kotrc, E. and Savicky, P. and Towers, S. and Vaiciulis, A. and Wittek, W.",
    title = "{Methods for multidimensional event classification: A case study using images from a Cherenkov gamma-ray telescope}",
    doi = "10.1016/j.nima.2003.08.157",
    journal = "Nucl. Instrum. Meth. A",
    volume = "516",
    pages = "511--528",
    year = "2004"
}

@article{Albert:2007yd,
    author = "Albert, J. and others",
    collaboration = "MAGIC",
    title = "{Implementation of the Random Forest Method for the Imaging Atmospheric Cherenkov Telescope MAGIC}",
    eprint = "0709.3719",
    archivePrefix = "arXiv",
    primaryClass = "astro-ph",
    doi = "10.1016/j.nima.2007.11.068",
    journal = "Nucl. Instrum. Meth. A",
    volume = "588",
    pages = "424--432",
    year = "2008"
}

@article{Ohm:2009nw,
    author = "Ohm, S. and van Eldik, C. and Egberts, K.",
    title = "{Gamma-Hadron Separation in Very-High-Energy gamma-ray astronomy using a multivariate analysis method}",
    eprint = "0904.1136",
    archivePrefix = "arXiv",
    primaryClass = "astro-ph.IM",
    doi = "10.1016/j.astropartphys.2009.04.001",
    journal = "Astropart. Phys.",
    volume = "31",
    pages = "383--391",
    year = "2009"
}

@article{Krause:2017erl,
    author = "Krause, Maria and Pueschel, Elisa and Maier, Gernot",
    title = "{Improved $\gamma$/hadron separation for the detection of faint $\gamma$-ray sources using boosted decision trees}",
    eprint = "1701.06928",
    archivePrefix = "arXiv",
    primaryClass = "astro-ph.IM",
    doi = "10.1016/j.astropartphys.2017.01.004",
    journal = "Astropart. Phys.",
    volume = "89",
    pages = "1--9",
    year = "2017"
}

@inproceedings{Holch:2017fqm,
    author = "Holch, Tim Lukas and Shilon, Idan and B{\"u}chele, Matthias and Fischer, Tobias and Funk, Stefan and Groeger, Nils and Jankowsky, David and Lohse, Thomas and Schwanke, Ullrich and Wagner, Philipp",
    title = "{Probing Convolutional Neural Networks for Event Reconstruction in $\gamma$-Ray Astronomy with Cherenkov Telescopes}",
    booktitle = "{35th International Cosmic Ray Conference}",
    eprint = "1711.06298",
    archivePrefix = "arXiv",
    primaryClass = "astro-ph.IM",
    doi = "10.22323/1.301.0795",
    journal = "PoS",
    volume = "ICRC2017",
    pages = "795",
    year = "2018"
}

@article{Shilon:2018xlp,
    author = "Shilon, Idan and Kraus, Manuel and B{\"u}chele, Matthias and Egberts, Kathrin and Fischer, Tobias and Holch, Tim Lukas and Lohse, Thomas and Schwanke, Ullrich and Steppa, Constantin and Funk, Stefan",
    title = "{Application of Deep Learning methods to analysis of Imaging Atmospheric Cherenkov Telescopes data}",
    eprint = "1803.10698",
    archivePrefix = "arXiv",
    primaryClass = "astro-ph.IM",
    doi = "10.1016/j.astropartphys.2018.10.003",
    journal = "Astropart. Phys.",
    volume = "105",
    pages = "44--53",
    year = "2019"
}

@inproceedings{Jacquemont:2021cfo,
    author = "Jacquemont, Mika{\"e}l and Vuillaume, Thomas and Benoit, Alexandre and Maurin, Gilles and Lambert, Patrick and Lamanna, Giovanni",
    title = "{First Full-Event Reconstruction from Imaging Atmospheric Cherenkov Telescope Real Data with Deep Learning}",
    booktitle = "{2021 International Conference on Content-Based Multimedia Indexing (CBMI)}",
    eprint = "2105.14927",
    archivePrefix = "arXiv",
    primaryClass = "astro-ph.IM",
    doi = "10.1109/CBMI50038.2021.9461918",
    year = "2021"
}

@inproceedings{Vuillaume:2021zmv,
    author = "Vuillaume, Thomas and Jacquemont, Mika{\"e}l and de Bony de Lavergne, Mathieu and Sanchez, David A. and Poireau, Vincent and Maurin, Gilles and Benoit, Alexandre and Lambert, Patrick and Lamanna, Giovanni",
    collaboration = "CTA LST Project",
    title = "{Analysis of the Cherenkov Telescope Array first Large-Sized Telescope real data using convolutional neural networks}",
    booktitle = "{37th International Cosmic Ray Conference}",
    eprint = "2108.04130",
    archivePrefix = "arXiv",
    primaryClass = "astro-ph.IM",
    doi = "10.22323/1.395.0703",
    journal = "PoS",
    volume = "ICRC2021",
    pages = "703",
    year = "2021"
}

@inproceedings{Miener:2025ixi,
    author = "Miener, Tjark and Viale, Ilaria and Lacave, Bastien and Cervi{\~n}o, Alexander",
    title = "{Deep Learning-Based Stereoscopic Event Reconstruction for CTAO using CTLearn}",
    booktitle = "{39th International Cosmic Ray Conference}",
    eprint = "2509.14973",
    archivePrefix = "arXiv",
    primaryClass = "astro-ph.IM",
    doi = "10.22323/1.501.0757",
    journal = "PoS",
    volume = "ICRC2025",
    pages = "757",
    year = "2025"
}

@article{Glombitza:2023qen,
    author = "Glombitza, Jonas and Joshi, Vikas and Bruno, Benedetta and Funk, Stefan",
    title = "{Application of graph networks to background rejection in Imaging Air Cherenkov Telescopes}",
    eprint = "2305.08674",
    archivePrefix = "arXiv",
    primaryClass = "astro-ph.IM",
    doi = "10.1088/1475-7516/2023/11/008",
    journal = "JCAP",
    volume = "11",
    pages = "008",
    year = "2023"
}

@article{Jobst:2026fuw,
    author = "Jobst, Elli and Heckmann, Lea and Heinrich, Lukas and Paneque, David",
    title = "{The Analysis, not the Aperture: End-to-End Transformer Reconstruction for Imaging Atmospheric Cherenkov Telescopes}",
    eprint = "2608.31148",
    archivePrefix = "arXiv",
    primaryClass = "astro-ph.IM",
    month = "8",
    year = "2026"
}

@article{Schwefer:2024bsx,
    author = "Schwefer, Georg and Parsons, Robert and Hinton, Jim",
    title = "{A hybrid approach to event reconstruction for atmospheric Cherenkov Telescopes combining machine learning and likelihood fitting}",
    eprint = "2406.17502",
    archivePrefix = "arXiv",
    primaryClass = "astro-ph.IM",
    doi = "10.1016/j.astropartphys.2024.103008",
    journal = "Astropart. Phys.",
    volume = "163",
    pages = "103008",
    year = "2024"
}

@article{Elflein:2023jct,
    author = "Elflein, Christian and Funk, Stefan and Glombitza, Jonas",
    title = "{Ultra-fast generation of air shower images for Imaging Air Cherenkov Telescopes using Generative Adversarial Networks}",
    eprint = "2311.01385",
    archivePrefix = "arXiv",
    primaryClass = "astro-ph.IM",
    doi = "10.1088/1748-0221/19/04/P04010",
    journal = "JINST",
    volume = "19",
    number = "04",
    pages = "P04010",
    year = "2024"
}

@article{Elflein:2026ycz,
    author = "Elflein, Christian and Funk, Stefan and Glombitza, Jonas and Mikuni, Vinicius and Nachman, Benjamin and Wang, Lark",
    title = "{Generation of Imaging Air Cherenkov Telescope images using Diffusion Models}",
    eprint = "2603.06364",
    archivePrefix = "arXiv",
    primaryClass = "astro-ph.IM",
    doi = "10.1088/1475-7516/2026/08/079",
    journal = "JCAP",
    volume = "08",
    pages = "079",
    year = "2026"
}

@article{MODE:2022znx,
    author = "Dorigo, Tommaso and others",
    collaboration = "MODE",
    title = "{Toward the end-to-end optimization of particle physics instruments with differentiable programming}",
    eprint = "2203.13818",
    archivePrefix = "arXiv",
    primaryClass = "physics.ins-det",
    doi = "10.1016/j.revip.2023.100085",
    journal = "Rev. Phys.",
    volume = "10",
    pages = "100085",
    year = "2023"
}

@article{Ullmo:2024odq,
    author = "Ullmo, Marion and Moulin, Emmanuel",
    title = "{Nonparametric signal separation in very-high-energy gamma ray observations with probabilistic neural networks}",
    eprint = "2407.01329",
    archivePrefix = "arXiv",
    primaryClass = "astro-ph.IM",
    doi = "10.1088/1475-7516/2025/01/014",
    journal = "JCAP",
    volume = "01",
    pages = "014",
    year = "2025"
}

@inproceedings{Hermans:2019ioj,
    author = "Hermans, Joeri and Begy, Volodimir and Louppe, Gilles",
    title = "{Likelihood-free MCMC with Amortized Approximate Ratio Estimators}",
    eprint = "1903.04057",
    archivePrefix = "arXiv",
    primaryClass = "stat.ML",
    booktitle = "{Proceedings of the 37th International Conference on Machine Learning (ICML 2020)}",
    year = "2020"
}

@article{Caron:2022akb,
    author = "Caron, Sascha and Eckner, Christopher and Hendriks, Luc and J{\'o}hannesson, Gu{\dh}laugur and Ruiz de Austri, Roberto and Zaharijas, Gabrijela",
    title = "{Mind the gap: the discrepancy between simulation and reality drives interpretations of the Galactic Center Excess}",
    eprint = "2211.09796",
    archivePrefix = "arXiv",
    primaryClass = "astro-ph.HE",
    doi = "10.1088/1475-7516/2023/06/013",
    journal = "JCAP",
    volume = "06",
    pages = "013",
    year = "2023"
}

@article{Mishra-Sharma:2021oxe,
    author = "Mishra-Sharma, Siddharth and Cranmer, Kyle",
    title = "{Neural simulation-based inference approach for characterizing the Galactic Center $\gamma$-ray excess}",
    eprint = "2110.06931",
    archivePrefix = "arXiv",
    primaryClass = "astro-ph.HE",
    doi = "10.1103/PhysRevD.105.063017",
    journal = "Phys. Rev. D",
    volume = "105",
    number = "6",
    pages = "063017",
    year = "2022"
}

@article{List:2021aer,
    author = "List, Florian and Rodd, Nicholas L. and Lewis, Geraint F.",
    title = "{Extracting the Galactic Center excess' source-count distribution with neural nets}",
    eprint = "2107.09070",
    archivePrefix = "arXiv",
    primaryClass = "astro-ph.HE",
    doi = "10.1103/PhysRevD.104.123022",
    journal = "Phys. Rev. D",
    volume = "104",
    number = "12",
    pages = "123022",
    year = "2021"
}

@article{Christy:2024gsl,
    author = "Christy, Katharena and Baxter, Eric J. and Kumar, Jason",
    title = "{Applying simulation-based inference to spectral and spatial information from the Galactic Center gamma-ray excess}",
    eprint = "2402.04549",
    archivePrefix = "arXiv",
    primaryClass = "astro-ph.HE",
    doi = "10.1088/1475-7516/2024/07/066",
    journal = "JCAP",
    volume = "07",
    pages = "066",
    year = "2024"
}

@inproceedings{AnauMontel:2022ppb,
    author = "Anau Montel, Noemi and Weniger, Christoph",
    title = "{Detection is truncation: studying source populations with truncated marginal neural ratio estimation}",
    booktitle = "{36th Conference on Neural Information Processing Systems: Workshop on Machine Learning and the Physical Sciences}",
    eprint = "2211.04291",
    archivePrefix = "arXiv",
    primaryClass = "astro-ph.IM",
    year = "2022"
}

@article{Bhattacharjee:2025ofd,
    author = "Bhattacharjee, Pooja and Eckner, Christopher and Zaharijas, Gabrijela and Kluge, Gert and D'Amico, Giacomo",
    title = "{Probing the Parameter Space of Axion-Like Particles Using Simulation-Based Inference}",
    eprint = "2509.20578",
    archivePrefix = "arXiv",
    primaryClass = "astro-ph.HE",
    month = "9",
    year = "2025"
}

@article{Dax:2021tsq,
    author = "Dax, Maximilian and Green, Stephen R. and Gair, Jonathan and Macke, Jakob H. and Buonanno, Alessandra and Sch{\"o}lkopf, Bernhard",
    title = "{Real-Time Gravitational Wave Science with Neural Posterior Estimation}",
    eprint = "2106.12594",
    archivePrefix = "arXiv",
    primaryClass = "gr-qc",
    doi = "10.1103/PhysRevLett.127.241103",
    journal = "Phys. Rev. Lett.",
    volume = "127",
    number = "24",
    pages = "241103",
    year = "2021"
}

@article{Dax:2024mcn,
    author = "Dax, Maximilian and Green, Stephen R. and Gair, Jonathan and Gupte, Nihar and P{\"u}rrer, Michael and Raymond, Vivien and Wildberger, Jonas and Macke, Jakob H. and Buonanno, Alessandra and Sch{\"o}lkopf, Bernhard",
    title = "{Real-time inference for binary neutron star mergers using machine learning}",
    eprint = "2407.09602",
    archivePrefix = "arXiv",
    primaryClass = "gr-qc",
    doi = "10.1038/s41586-025-08593-z",
    journal = "Nature",
    volume = "639",
    number = "8053",
    pages = "49--53",
    year = "2025"
}

@article{Jeffrey:2020xve,
    author = "Jeffrey, Niall and Alsing, Justin and Lanusse, Fran{\c{c}}ois",
    title = "{Likelihood-free inference with neural compression of DES SV weak lensing map statistics}",
    eprint = "2009.08459",
    archivePrefix = "arXiv",
    primaryClass = "astro-ph.CO",
    doi = "10.1093/mnras/staa3594",
    journal = "Mon. Not. Roy. Astron. Soc.",
    volume = "501",
    number = "1",
    pages = "954--969",
    year = "2021"
}

@article{DES:2024xij,
    author = "Jeffrey, N. and others",
    collaboration = "DES",
    title = "{Dark Energy Survey Year 3 results: likelihood-free, simulation-based $w$CDM inference with neural compression of weak-lensing map statistics}",
    eprint = "2403.02314",
    archivePrefix = "arXiv",
    primaryClass = "astro-ph.CO",
    doi = "10.1093/mnras/stae2629",
    journal = "Mon. Not. Roy. Astron. Soc.",
    volume = "536",
    number = "2",
    pages = "1303--1322",
    year = "2024"
}

@article{Hahn:2022zxa,
    author = "Hahn, ChangHoon and Eickenberg, Michael and Ho, Shirley and Hou, Jiamin and Lemos, Pablo and Massara, Elena and Modi, Chirag and Moradinezhad Dizgah, Azadeh and R{\'e}galdo-Saint Blancard, Bruno and Abidi, Muntazir M.",
    title = "{SIMBIG: A Forward Modeling Approach To Analyzing Galaxy Clustering}",
    eprint = "2211.00723",
    archivePrefix = "arXiv",
    primaryClass = "astro-ph.CO",
    doi = "10.1073/pnas.2218810120",
    journal = "Proc. Nat. Acad. Sci.",
    volume = "120",
    number = "42",
    pages = "e2218810120",
    year = "2023"
}

@article{ATLAS:2024jry,
    author = "Aad, Georges and others",
    collaboration = "ATLAS",
    title = "{Measurement of off-shell Higgs boson production in the $H^*\rightarrow ZZ\rightarrow 4\ell$ decay channel using a neural simulation-based inference technique in 13 TeV $pp$ collisions with the ATLAS detector}",
    eprint = "2412.01548",
    archivePrefix = "arXiv",
    primaryClass = "hep-ex",
    doi = "10.1088/1361-6633/adcd9a",
    journal = "Rept. Prog. Phys.",
    volume = "88",
    number = "5",
    pages = "057803",
    year = "2025"
}

@article{HESS:2005MSH,
    author = "Aharonian, F. and others",
    collaboration = "H.E.S.S.",
    title = "{Discovery of extended VHE gamma-ray emission from the asymmetric pulsar wind nebula in MSH 15-52 with H.E.S.S.}",
    eprint = "astro-ph/0504120",
    archivePrefix = "arXiv",
    doi = "10.1051/0004-6361:200500105",
    journal = "Astron. Astrophys.",
    volume = "435",
    pages = "L17",
    year = "2005"
}

@article{Knodlseder:2019ctools,
    author = {Kn{\"o}dlseder, J{\"u}rgen and Tibaldo, Luigi and Tiziani, Domenico and Specovius, Andreas and Cardenzana, Josh and others},
    title = "{Analysis of the H.E.S.S. public data release with ctools}",
    eprint = "1910.09456",
    archivePrefix = "arXiv",
    primaryClass = "astro-ph.IM",
    journal = "Astron. Astrophys.",
    volume = "632",
    pages = "A102",
    year = "2019"
}

@article{HESS:2018HGPS,
    author = "Abdalla, H. and others",
    collaboration = "H.E.S.S.",
    title = "{The H.E.S.S. Galactic plane survey}",
    eprint = "1804.02432",
    archivePrefix = "arXiv",
    primaryClass = "astro-ph.HE",
    doi = "10.1051/0004-6361/201732098",
    journal = "Astron. Astrophys.",
    volume = "612",
    pages = "A1",
    year = "2018"
}

@article{Eckner:2022gil,
    author = "Eckner, Christopher and Vodeb, Veronika and Martin, Pierrick and Zaharijas, Gabrijela and Calore, Francesca",
    title = "{Detecting and characterizing pulsar haloes with the Cherenkov telescope array}",
    eprint = "2212.11265",
    archivePrefix = "arXiv",
    primaryClass = "astro-ph.HE",
    reportNumber = "LAPTH-071/22",
    doi = "10.1093/mnras/stad715",
    journal = "Mon. Not. Roy. Astron. Soc.",
    volume = "521",
    number = "3",
    pages = "3793--3809",
    year = "2023"
}

@article{Mestre:2022kip,
    author = "Mestre, Enrique and Torres, Diego F. and de Ona Wilhelmi, Emma and Mart{\'\i}, Josep",
    title = "{Testing source confusion and identification capability in Cherenkov telescope array data}",
    eprint = "2210.04344",
    archivePrefix = "arXiv",
    primaryClass = "astro-ph.HE",
    doi = "10.1093/mnras/stac2910",
    journal = "Mon. Not. Roy. Astron. Soc.",
    volume = "517",
    number = "3",
    pages = "3550--3567",
    year = "2022"
}

@article{HAWC:2017kbo,
    author = "Abeysekara, A. U. and others",
    collaboration = "HAWC",
    title = "{Extended gamma-ray sources around pulsars constrain the origin of the positron flux at Earth}",
    eprint = "1711.06223",
    archivePrefix = "arXiv",
    primaryClass = "astro-ph.HE",
    doi = "10.1126/science.aan4880",
    journal = "Science",
    volume = "358",
    number = "6365",
    pages = "911--914",
    year = "2017"
}

@article{Bose:2021oiq,
    author = "Bose, D. and Chitnis, V. R. and Majumdar, P. and Acharya, B. S.",
    title = "{Ground-based gamma-ray astronomy: history and development of techniques}",
    eprint = "2201.04719",
    archivePrefix = "arXiv",
    primaryClass = "astro-ph.IM",
    doi = "10.1140/epjs/s11734-021-00396-3",
    journal = "Eur. Phys. J. ST",
    volume = "231",
    number = "1",
    pages = "3--26",
    year = "2022"
}

@inbook{Holder:GroundbasedGammaRays,
author = {Jamie Holder},
title = {Atmospheric Cherenkov Gamma-Ray Telescopes},
booktitle = {The WSPC Handbook of Astronomical Instrumentation},
chapter = {Chapter 6},
pages = {117-136},
year = "2022",
doi = {10.1142/9789811203817_0006},
URL = {https://www.worldscientific.com/doi/abs/10.1142/9789811203817_0006},
eprint = {https://www.worldscientific.com/doi/pdf/10.1142/9789811203817_0006},
}

@article{deNaurois:2009ud,
    author = "de Naurois, Mathieu and Rolland, Loic",
    title = "{A high performance likelihood reconstruction of gamma-rays for Imaging Atmospheric Cherenkov Telescopes}",
    eprint = "0907.2610",
    archivePrefix = "arXiv",
    primaryClass = "astro-ph.IM",
    doi = "10.1016/j.astropartphys.2009.09.001",
    journal = "Astropart. Phys.",
    volume = "32",
    pages = "231",
    year = "2009"
}

@article{CTA-LSTProject:2023haa,
    author = "Abe, H. and others",
    collaboration = "CTA-LST Project",
    title = "{Observations of the Crab Nebula and Pulsar with the Large-sized Telescope Prototype of the Cherenkov Telescope Array}",
    eprint = "2306.12960",
    archivePrefix = "arXiv",
    primaryClass = "astro-ph.HE",
    doi = "10.3847/1538-4357/ace89d",
    journal = "Astrophys. J.",
    volume = "956",
    number = "2",
    pages = "80",
    year = "2023"
}

@article{Deil:2016yrp,
    author = "Deil, Christoph and others",
    title = "{Open high-level data formats and software for gamma-ray astronomy}",
    eprint = "1610.01884",
    archivePrefix = "arXiv",
    primaryClass = "astro-ph.IM",
    doi = "10.1063/1.4969003",
    journal = "AIP Conf. Proc.",
    volume = "1792",
    number = "1",
    pages = "070006",
    year = "2017"
}

\appendix

\section{Tests of additional systematic modes}
\label{app:modes}

We tested whether the five modes of Eq.~\eqref{eq:basis} miss a deviation that recurs in off-field observations without known gamma-ray sources. The candidate is a coupling of the radial and the energy dependence, $(r/R_0-1)(\mathcal{E}-\bar{\mathcal{E}})/\sigma_\mathcal{E}$, motivated by the energy dependence of the offset acceptance (Sec.~\ref{sec:anisotropy}); as a control, we used the physically unmotivated term $xy/R_0^2$. For each of the 55 pool runs, we binned the observed counts and the prediction of its owner ensemble in cells of $0.2^\circ\times0.2^\circ$ and six energy bins between 0.5 and 20\,TeV and fitted the five modes plus the candidate by Poisson regression. Before the fit, we required for a detection a coefficient of at least 0.05 in absolute value, a 99\% whole-run bootstrap interval that excludes zero, and a control coefficient consistent with zero. The candidate coefficient is $+0.025$, with a 99\% interval of $[-0.009,+0.059]$; the control coefficient is $-0.001$. Neither term is required at the sensitivity of the release. We, therefore, keep the five-mode basis.

\begin{figure}[h]
    \centering
    \includegraphics[width=0.9\columnwidth]{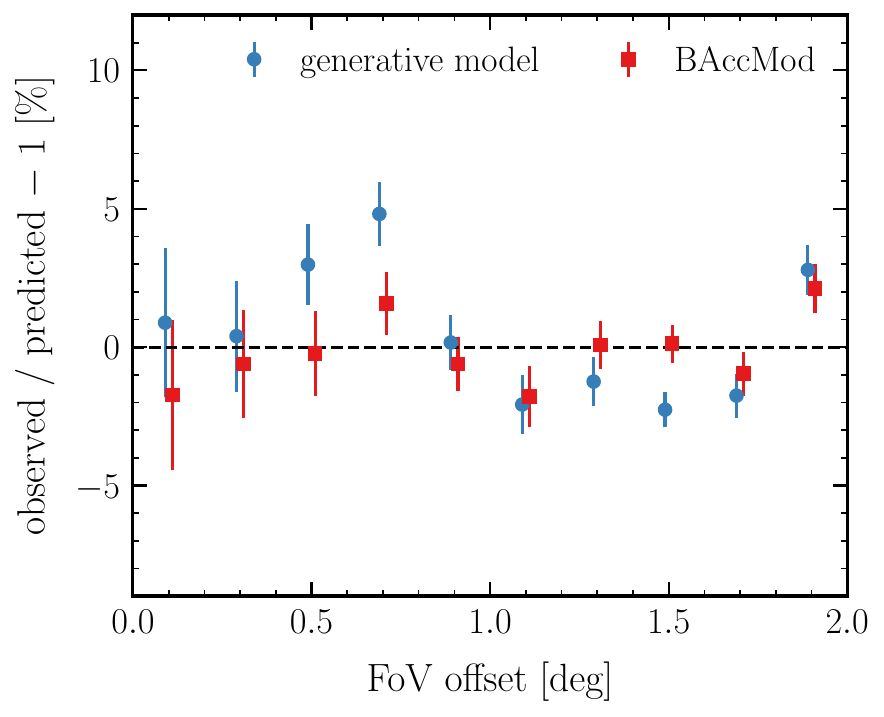}
    \caption{Radial residuals of the stacked cross-fitted predictions for the 55 pool runs, after a single normalization of the stack, in rings of $0.2^\circ$ between $\max(0.5\,\mathrm{TeV},E_\mathrm{safe})$ and 20\,TeV. Error bars are standard deviations from bootstrapping whole runs; the BAccMod points are the best setup of Appendix~\ref{app:baccmod}, shifted slightly to the right for visibility.}
    \label{fig:radial}
\end{figure}

The stack of all 55 pool runs, however, reveals a radial structure that is common to the runs (Fig.~\ref{fig:radial}). After a single normalization of the stack, the cross-fitted prediction falls short by up to 5\% at offsets of 0.4--0.8$^\circ$ and exceeds the data by about 2\% at 1.0--1.8$^\circ$; with run-bootstrap errors per ring of $0.2^\circ$, this corresponds to $\chi^2=54$ for 9 degrees of freedom. The radial mode, being linear in the offset, absorbs little of it ($\chi^2=47$ for 8 degrees of freedom) and, in fact, the systematic handler is not designed to do so: it describes zero-mean deviations of individual runs, whereas this structure is a deviation of the mean intensity that does not average down in a stack. The radial model of BAccMod, derived from the same runs, follows the same pattern at a lower amplitude that is not significant on its own ($\chi^2=13$). Since two methods with entirely different constructions share the pattern, we conjecture that it reflects the limited livetime and number of runs of the release rather than a deficiency specific to the flow and that a larger training sample will resolve it.

\section{Validation of neural posteriors}
\label{app:sbi-val}

The simulator defines a joint distribution of parameters and data based on the prior $p(\bm\theta)$ and the likelihood $p(\bm x\,|\,\bm\theta)$ that it samples implicitly. Neural posterior estimation returns an approximation (or estimator) $q(\bm\theta\,|\,\bm x)$ to the true posterior $p(\bm\theta\,|\,\bm x)\propto p(\bm x\,|\,\bm\theta)\,p(\bm\theta)$ for any $\bm x$, without retraining. Since the true posterior is unknown or at least inaccessible for all practical purposes, $q$ cannot be compared with it directly. Its quality can, however, be tested on simulations with known parameters. For this purpose, we draw $N=1000$ pairs $(\bm\theta_i,\bm x_i)$ from the prior and the simulator, including the Poisson noise and a new deviation from the systematic handler $\bm z_k$ for every run. These samples are a separate calibration dataset that played no role in training or in the selection of the epoch. For each calibration simulation, we draw $L=2048$ samples $\bm\theta_i^{(l)}\sim q(\bm\theta\,|\,\bm x_i)$ from the respective posterior. Such tests establish that $q$ is faithful \emph{on average over the prior} \citep{Talts:2018zdk, Hermans:2021rqv}; they cannot tell whether the simulator describes the observed data, i.e., identify model mis-specification. The latter is the task of the posterior predictive check. We apply the diagnostics below to every posterior in Sec.~\ref{sec:results}.

\paragraph{Simulation-based calibration.} For every parameter $\theta_j$ and test simulation $i$, the normalized rank of the true value among the posterior draws is
\begin{equation}
    u_{ij}=\frac1L\sum_{l=1}^{L}\mathbbm{1}\!\left[\theta_{ij}^{(l)}<\theta_{ij}\right]\in[0,1].
    \label{eq:sbc-rank}
\end{equation}
If $q$ equals the true posterior, $\theta_{ij}$ and the draws $\bm\theta_i^{(l)}\sim q(\bm\theta\,|\,\bm x_i)$ are independent samples of the same distribution, so that $u_{ij}$ is uniformly distributed over the $N$ test simulations \citep{Cook:2006, Talts:2018zdk}. The shape of the rank histogram is an indicator of how the posterior may fail: a U-shape signals posteriors that are too narrow, inverse U-shape posteriors  occur when they are too wide, and a slope hints at a bias. We test uniformity per parameter with a Kolmogorov--Smirnov test. Simulation-based calibration is only a necessary condition that tests each marginal separately. It is passed by any posterior that is calibrated on average, which also includes  the prior itself, that is, $q(\bm\theta\,|\,\bm x)=p(\bm\theta)$.

\paragraph{Expected coverage of joint credible regions.} The highest-posterior-density (HPD) region of credibility $c$ is the smallest region that contains the probability $c$,
\begin{equation}
    \Theta_c(\bm x)=\{\bm\theta:\,q(\bm\theta\,|\,\bm x)\geq t_c(\bm x)\},\qquad \int_{\Theta_c(\bm x)}q(\bm\theta\,|\,\bm x)\,\mathrm d\bm\theta=c
\end{equation}
\citep{Hyndman:1996}. A calibrated posterior satisfies $P[\bm\theta\in\Theta_c(\bm x)]=c$ for every $c$, where the probability refers to the joint distribution of $\bm\theta$ and $\bm x$. Because a flow provides the normalized density $q$, membership follows from the posterior mass above the truth,
\begin{equation}
    \alpha_i=\frac1L\sum_{l=1}^{L}\mathbbm{1}\!\left[q(\bm\theta_i^{(l)}\,|\,\bm x_i)>q(\bm\theta_i\,|\,\bm x_i)\right],
\end{equation}
since $\bm\theta_i\in\Theta_c(\bm x_i)$ if and only if $\alpha_i\leq c$. The empirical expected coverage is thus $\hat C(c)=N^{-1}\sum_i\mathbbm{1}[\alpha_i\leq c]$ \citep{Hermans:2021rqv}. $\hat C(c)>c$ indicates conservative, $\hat C(c)<c$ overconfident posteriors. In contrast to the ranks, the test acts on the full joint density and therefore also checks the correlations between the parameters, such as that between $N_0$ and $\eta$. We express credibility in units of Gaussian standard deviations, $z(c)=\Phi^{-1}[(1+c)/2]$ with the standard normal distribution function $\Phi$, so that $c=0.683$ and $0.954$ correspond to $z=1$ and 2, and compare the empirical $z[\hat C(c)]$ with the nominal $z(c)$ for $0\leq z(c)\leq3.5$. The number of covered simulations is binomial; we show its 68\% Jeffreys interval, i.e.\ the central interval of a $\mathrm{Beta}(n+\tfrac12,N-n+\tfrac12)$ distribution for $n$ covered simulations \citep{Brown:2001}, converted to $z$ in the same way. With $N=1000$, coverages beyond about $3.3\sigma$ ($c=1-1/N$) are not resolved, so that full containment of all test simulations is a lower bound rather than a measurement. We report the largest deviation $|\Delta z|$ between empirical and nominal coverage on this domain.

\paragraph{Informativeness and prior boundaries.} A calibrated posterior may still carry little information. For every parameter, we compare the central 68\% interval of the posterior of the observed data, of width $w_j$, with that of the prior, which for a uniform prior on $[a_j,b_j]$ is $0.68\,(b_j-a_j)$, through the contraction $\kappa_j=1-w_j/[0.68\,(b_j-a_j)]$. A contraction near zero means that the data barely constrain $\theta_j$, one near unity a posterior much narrower than the prior. The prior ranges must moreover not cut the posterior, which would make calibration an artifact of the prior. We therefore record the fraction of posterior draws in the outer 2\% of each prior range.

\paragraph{Posterior predictive check.} The previous tests use only simulations, which cannot tell us anything about model mis-specification by construction. A simulator that fails to describe the observed data $\bm x_\mathrm{obs}$ can pass all previous checks. The posterior predictive distribution
\begin{equation}
    p(\tilde{\bm x}\,|\,\bm x_\mathrm{obs})=\int p(\tilde{\bm x}\,|\,\bm\theta)\,q(\bm\theta\,|\,\bm x_\mathrm{obs})\,\mathrm d\bm\theta
\end{equation}
describes the dataset that the fitted model expects in a replication of the observation \citep{Rubin:1984, Gelman:1996}. We draw $M=512$ parameter vectors from $q(\bm\theta\,|\,\bm x_\mathrm{obs})$, simulate one cube $\tilde{\bm x}^{(m)}$ for each, with new run deviations $\bm z_k$ of the systematic handler and Poisson noise, and compare a suitable test statistic $T$ of the replicas with its observed value through the two-sided predictive $p$-value
\begin{equation}
    p=\min\!\left\{1,\;2\min\!\left(\hat P[T(\tilde{\bm x})\leq T(\bm x_\mathrm{obs})],\;\hat P[T(\tilde{\bm x})\geq T(\bm x_\mathrm{obs})]\right)\right\}.
\end{equation}
As a simple choice for $T$, we adopt the total counts in the energy band of the analysis. The power of such a check depends on the width of the predictive distribution. A distribution that is much wider than the uncertainty on $T$ that matters for the analysis accepts almost any observation, and a very narrow one rejects even small imperfections of the simulator. We therefore quote with every $p$-value the relative width $r=\mathrm{sd}[T(\tilde{\bm x})]/T(\bm x_\mathrm{obs})$. Since the replicas redraw the run deviations $\bm z_k$, $r$ is dominated by the systematic handler rather than by Poisson noise and a check with relative width $r$ can only detect a mismatch of the total counts of order $2r$ or larger.

\begin{figure}[h]
    \centering
    \includegraphics[width=0.95\columnwidth]{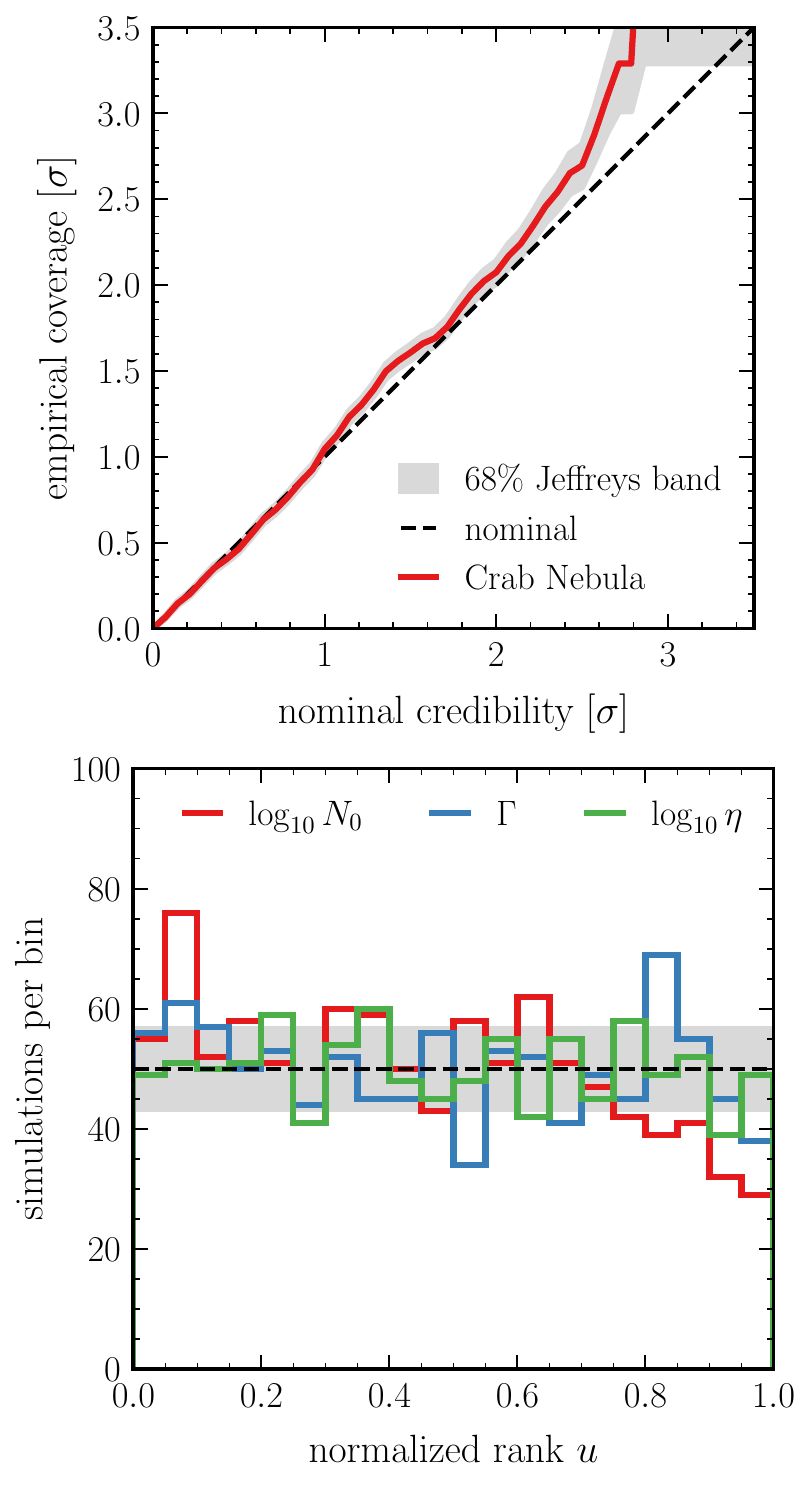}
    \caption{Calibration of the posterior of the Crab Nebula on 1000 calibration samples. Top: expected coverage of joint HPD regions against their nominal credibility, both in units of Gaussian standard deviations, with the 68\% Jeffreys interval of the binomial count; the dashed line marks exact calibration. Bottom: histograms of the normalized ranks $u$ (Eq.~\ref{eq:sbc-rank}) of the true parameters in 20 bins; the dashed line and the gray band show the expectation for uniform ranks and its central 68\% range.}
    \label{fig:crab-calib}
\end{figure}

\paragraph{Acceptance criteria.} Before the analyses, we fixed the conditions under which we quote a posterior: a contraction $\kappa_j\geq0.2$ and an occupancy of the outer 2\% of the prior range below 1\% on both sides for every parameter, a Kolmogorov--Smirnov $p$-value above 0.05 for the ranks of every parameter, and $|\Delta z|\leq2$ for the joint coverage. For the Crab Nebula (Sec.~\ref{sec:results}), the contractions are 0.98, 0.93 and 0.81 for $\log_{10}N_0$, $\Gamma$ and $\log_{10}\eta$, the prior edges are occupied by at most $5\times10^{-5}$ of the draws, and $|\Delta z|=0.64$ (Fig.~\ref{fig:crab-calib}). The ranks of $\Gamma$ and $\log_{10}\eta$ are uniform ($p=0.33$ and 0.92), but those of $\log_{10}N_0$ are not ($p=2\times10^{-5}$), so that the posterior of the Crab Nebula fails one of our criteria. The rank histogram of $\log_{10}N_0$ falls from low to high ranks: the true value lies, on average, $0.16\pm0.03$ posterior standard deviations below the posterior median, as measured by the mean of $\Phi^{-1}(u)$, while the width of the posterior is correct. The offset stems from the simulations with true indices $\Gamma<2$ and $\Gamma\geq3$ ($0.22$ standard deviations each, $p=8\times10^{-4}$). For $2\leq\Gamma<3$, which contains the posterior of the Crab Nebula, the ranks are uniform ($p=0.54$), with an offset of $0.02\pm0.05$ standard deviations, i.e.\ at most 0.4\% in $N_0$ for the width of the Crab posterior. We therefore quote the posterior, noting that the calibration of $N_0$ holds only locally. 

\section{BAccMod setups and comparison over the pool}
\label{app:baccmod}

We ran BAccMod 0.4.1 in the cross-fitting scheme of Sec.~\ref{sec:bkg-val} with eight setups (Table~\ref{tab:baccmod}). All models extend to $2.5^\circ$ and are evaluated within $R_\mathrm{FoV}=2^\circ$ in the FoV frame of our model, with the livetime as exposure. For numerical reasons, we had to introduce two constraints on the analyses. First, the models start at 0.4\,TeV, where runs at all zenith angles still contain events, since empty energy bins drive BAccMod's logarithmic interpolation of its merged energy bins to zero or divergence. Second, we hold every model constant beyond its outermost spatial bin centers instead of extrapolating its logarithm, which diverges next to empty bins; this changes regular predictions by less than $10^{-4}$.

The 3D stack diverges for several runs even with these numerical safeguard rules, since the pool is too sparse for a histogram in two FoV coordinates and energy per zenith bin, and the 3D fit with a Gaussian times linear gradients does not describe the H.E.S.S.\ acceptance. The three radial setups perform equally well; we select the zenith window, the setup with the narrowest significance distribution after run-wise rescaling. Without any zenith dependence, the shape is equally good, but the normalization of individual runs scatters by 65\%. Over the 55 pool runs, the cross-fitted significance distributions have widths of 1.043 for the generative model and 1.063 for BAccMod, a paired difference of $0.020\pm0.014$ from run bootstrapping, which persists after run-wise rescaling ($0.023\pm0.013$). The normalization of individual runs scatters by 15\% and 16\%, respectively.

\begin{table}[h]
\centering
\small
\setlength{\tabcolsep}{3pt}
\begin{tabular}{lrrrr}
\toprule
setup & $\mu$ & $\sigma$ & $\sigma_\mathrm{resc}$ & scatter \\
\midrule
generative model & $-0.06$ & 1.043 & 1.046 & 15\% \\
radial, zenith window $\pm5^\circ$ & $-0.31$ & 1.063 & 1.069 & 16\% \\
radial, zenith bins $\geq2$\,h & $-0.31$ & 1.066 & 1.070 & 16\% \\
radial, zenith bins $\geq4$\,h & $-0.24$ & 1.069 & 1.070 & 16\% \\
radial, no zenith dependence & $-0.98$ & 1.060 & 1.071 & 65\% \\
3D fit, zenith bins $\geq2$\,h & $-1.41$ & 1.523 & 1.399 & 21\% \\
3D fit, zenith bins $\geq4$\,h & $-1.45$ & 1.495 & 1.390 & 16\% \\
3D stack, zenith bins $\geq2$\,h & \multicolumn{4}{c}{diverges for 46 runs} \\
3D stack, zenith bins $\geq4$\,h & \multicolumn{4}{c}{diverges for 6 runs} \\
\bottomrule
\end{tabular}
\caption{Cross-fitted predictions for the 55 pool runs: mean $\mu$ and width $\sigma$ of the significance distribution, width $\sigma_\mathrm{resc}$ after rescaling every run to its own counts, and scatter of the observed-to-predicted counts of individual runs. Zenith bins are formed in $\cos\theta_\mathrm{z}$ with a minimum livetime and interpolated; the zenith window selects the runs within $\pm5^\circ$ of the target run and uses fixed energy and offset bins.}
\label{tab:baccmod}
\end{table}

\section{Further checks of the Crab field}
\label{app:crab}

The background offset of the Crab field is field-wide: all four runs exceed the prediction, by 7--18\%, and omitting any single run changes the stacked offset by at most 1.7 percentage points. A regression of the residuals of the off runs on their recorded zenith angle and energy threshold predicts residuals of $-2$ to $-3\%$ for the Crab runs, the opposite sign. The deterministic grid prediction does not depend on the importance sample. Finally, the coefficient of the coupling between the radial and the energy dependence (Appendix~\ref{app:modes}) is $-0.17$ for the Crab runs, with a 95\% interval of $[-0.20,-0.13]$, far from the value of the training pool; the residual of the Crab field is therefore not the imprint of a common deficiency of the model.

\section{Additional validation tests for MSH~15$-$52}
\label{app:msh-val}

\begin{figure}[h]
    \centering
    \includegraphics[width=0.95\columnwidth]{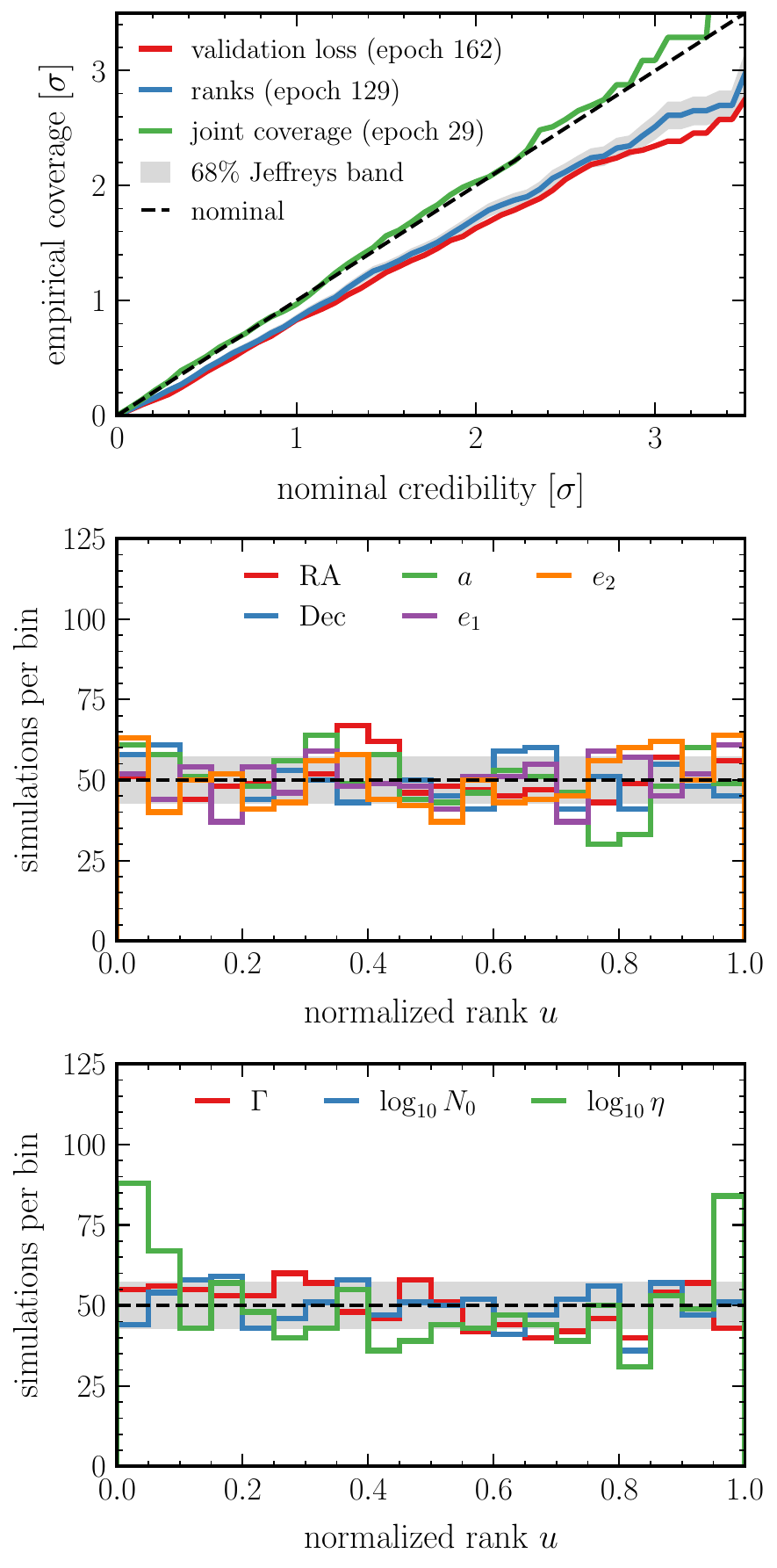}
    \caption{Calibration of the MSH~15$-$52 posteriors with the benchmark generative background model on 1000 `unseen' simulations. Top: expected coverage of joint HPD regions for the epochs of one training run selected by the validation loss (epoch 162), the marginal ranks (epoch 129, used in Sec.~\ref{sec:result-msh}) and the joint coverage (epoch 29), with the 68\% Jeffreys interval of the epoch-129 network. Middle and bottom: rank histograms of the epoch-129 network in 20 bins; the dashed line and the gray band show the expectation for uniform ranks and its central 68\% range.}
    \label{fig:msh-calib}
\end{figure}

The analysis of Sec.~\ref{sec:result-msh} uses the epoch with the most uniform marginal ranks. To understand how the calibration of the posterior evolves during training, we additionally stored the epoch with the lowest validation loss and the epoch with the best joint coverage. The two calibration metrics were measured after every epoch of the same training run on a separate store of 2000 simulations. The metric to beat was defined as the mean absolute deviation of the joint HPD coverage from the diagonal between 0 and $3\sigma$ and by the largest Kolmogorov--Smirnov distance of the marginal ranks from uniformity. All three networks were then evaluated on the same 1000 independent simulations used in Sec.~\ref{sec:result-msh} (Fig.~\ref{fig:msh-calib}).

The three selections reveal two distinct regimes that occur during training. We characterize the marginals by the mean and the standard deviation of the probit-transformed ranks $\Phi^{-1}(u_{ij})$, which are 0 and 1 for a calibrated marginal; a mean offset indicates a bias in units of the posterior width, and a standard deviation $s>1$ ($s<1$) a marginal that is too narrow (too wide), by roughly $|s-1|$. Late in the training, the validation loss and the marginal ranks reach their optima at epochs 162 and 129. Both networks yield an overconfident joint posterior: its coverage falls short of the nominal levels by $0.29\sigma$ and $0.24\sigma$ on average between 0 and $3\sigma$. At epoch 162, the marginals are offset by up to 0.17 ($\log_{10}N_0$), and the ranks of six of the eight parameters fail the test (largest rank distance 0.071). At epoch 129, the offsets are at most 0.07 and the widths of all marginals are within 8\% of calibration except for $\log_{10}\eta$, whose marginal is 21\% too narrow and which determines the largest rank distance of 0.063. Early in the training, at epoch 29, the joint coverage is consistent with calibration (mean deviation of $0.04\sigma$ from the nominal levels), but the marginals are offset by up to 0.40 ($\log_{10}N_0$; 0.24 for $a$ and $e_2$), and the marginal of $\log_{10}N_0$ is 21\% too wide; the largest rank distance is 0.21. The joint test is passed here because the shifted marginals are compensated by their width. The networks trained with the field-trained background of Appendix~\ref{app:msh-infield} behave in the same way. Joint coverage and marginal ranks thus probe different properties of the posterior, which are not simultaneously met in this example: the ranks are blind to the dependence between the parameters, and the joint coverage does not see biases that are compensated by width. The validation loss, finally, is dominated by the bulk of the posterior and hardly reacts to the calibration of its tails, so that it continues to decrease (indicating that overfitting the training data is probably not the issue) while the joint posterior becomes overconfident.

The background normalization $\eta$ is the only parameter whose ranks fail clearly but the degree of failure is different for two regimes. The first regime is an edge effect. When the true $\eta$ lies close to the boundary of its prior, the posterior should be cut off sharply there. The flow cannot produce such a sharp cut and instead shifts the posterior towards the middle of the prior, so that $\eta$ comes out biased in these simulations. Our prior on $\eta$ is narrow ($\pm0.1$\,dex) compared with an earlier choice ($\pm0.25$\,dex), so that a larger share of the calibration simulations falls close to an edge (about a quarter instead of a tenth). This makes the effect more visible in the rank test, although the network itself does not behave worse. The second regime remains away from the edges: there, the posterior of $\eta$ is mildly too narrow. For MSH~15$-$52, only this second part matters because the inferred $\eta$ lies well inside the prior.

For the present proof of concept, we report the rank selection, whose miscalibration is essentially confined to $\eta$ and a moderately overconfident joint posterior, and note that the posterior of Sec.~\ref{sec:result-msh} does not pass all criteria of Appendix~\ref{app:sbi-val}. Several established remedies can be expected to improve on this. Larger training sets are the most direct one, since $2\times10^4$ simulations are sufficient but also modest for an eight-dimensional posterior with a free morphology. Ensembles of independently trained flows, whose averaged densities are known to be better calibrated and more conservative than their members \citep{Lakshminarayanan:2017, Hermans:2021rqv}, address the overconfidence of the joint posterior. Training objectives that explicitly penalize overconfidence, by a balancing condition \citep{Delaunoy:2022, Delaunoy:2023} or by a differentiable coverage term \citep{Falkiewicz:2023}, act on the same deficiency during training. A prior without hard edges for $\log_{10}\eta$, e.g.\ a Gaussian, avoids the truncation at the prior boundaries in the first place. Finally, the epoch can be selected by a criterion that requires both the joint coverage and the marginal ranks to be acceptable on an independent store.

\section{In-field analysis of MSH~15$-$52}
\label{app:msh-infield}

The runs of MSH~15$-$52 contain enough source-free area to derive the background from the field itself. We analyzed the field in two such configurations. First, a generative model trained only on the 20 MSH runs, where we exclude the source within $0.4^\circ$ and cross-fit in five repeats of five folds, so that every run is predicted by five members trained on 16 other runs of the field (as for the benchmark generative CR background model). Second, BAccMod with its 3D fit method (a two-dimensional Gaussian times linear gradients in the FoV), built from the 20 runs without zenith binning, both in-sample and cross-fitted on the same folds. For the background tests, all predictors were evaluated between 0.5 and 10\,TeV, the upper end of the BAccMod models; the classical fits use per-run counts cubes with $0.02^\circ$ pixels in the same band and a free background normalization per run, whereas the SBI analysis is based on data cubes with a spatial binning of $0.04^\circ$.

Table~\ref{tab:msh-infield-bkg} summarizes the background tests. With one free normalization per run and a correlation radius of $0.04^\circ$, as customary for FoV-background analyses, all predictors describe the field well. In the absolute test of Sec.~\ref{sec:result-bkg}, the field-trained generative model predicts the stacked counts to $+0.3\%$ with a width of 1.04, whereas the width for the BAccMod models is 1.15, a difference of $0.110\pm0.016$ after scaling every run to its counts (paired whole-run bootstrap). One run (obs.\ 20283) has 40--45\% fewer counts than predicted by every model, which points to a peculiar run rather than an issue of the background models.

\begin{table}[h]
\centering
\small
\setlength{\tabcolsep}{3pt}
\begin{tabular}{lrrrrr}
\toprule
 & \multicolumn{2}{c}{FoV norm} & \multicolumn{3}{c}{absolute} \\
background & $\mu$ & $\sigma$ & $\mu$ & $\sigma$ & bias \\
\midrule
generative, field & $-0.03$ & 1.01 & $-0.02$ & 1.04 & $+0.3\%$ \\
generative, benchmark & $-0.04$ & 1.02 & $-0.77$ & 1.07 & $-9.6\%$ \\
BAccMod, in-sample & $-0.10$ & 1.03 & $-0.18$ & 1.15 & $-1.3\%$ \\
BAccMod, cross-fitted & $-0.10$ & 1.03 & $-0.19$ & 1.15 & $-1.3\%$ \\
\bottomrule
\end{tabular}
\caption{Background tests for the 20 MSH runs between 0.5 and 10\,TeV. FoV norm: significance outside the source exclusion with one free normalization per run, $0.04^\circ$ correlation radius and $0.02^\circ$ pixels. Absolute: known-background significance of Sec.~\ref{sec:result-bkg} ($0.06^\circ$ top-hat) and stacked bias (observed/predicted $-1$) of the predictions as they are.}
\label{tab:msh-infield-bkg}
\end{table}

Table~\ref{tab:msh-infield-fit} lists the source parameters. The classical fits with the three backgrounds agree in morphology and with \citet{Mohrmann:2019hfq}, and their spectral indices of 2.16--2.27 are harder than what we found when fitting up to 20\,TeV (see Table~\ref{tab:msh}). The SBI analysis with the field-trained background (rank selection, epoch 117; otherwise as in Sec.~\ref{sec:result-msh}) gives RA $=228.551^\circ\pm0.012^\circ$, Dec $=-59.175^\circ\pm0.007^\circ$, $a=0.185^\circ\pm0.015^\circ$, $e=0.75^{+0.07}_{-0.09}$, $\phi=155^{+8}_{-9}$\,deg, $\Gamma=2.40\pm0.07$, $N_0=(3.27^{+0.20}_{-0.19})\times10^{-12}\,\mathrm{cm^{-2}\,s^{-1}\,TeV^{-1}}$ and $\eta=1.06\pm0.04$, consistent with the analysis of Sec.~\ref{sec:result-msh} within one standard deviation in every source parameter and, through $\eta\simeq1$, with the background test above. Its calibration is somewhat better than in the benchmark case: only the ranks of $\Gamma$ and $\log_{10}\eta$ are not uniform, and only marginally ($p=0.031$ and 0.020); the marginal of $\log_{10}\eta$ is 17\% too narrow, the others at most 7\%; and the HPD regions cover on average $0.19\sigma$ less than their nominal levels. The observed total of counts lies within the posterior predictive distribution ($p=0.36$, relative width 0.057). The residuals, with the background deformations refitted as in Sec.~\ref{sec:result-msh}, resemble those with the benchmark model (mean $-0.09$, standard deviation 1.06 for the smoothed map), including the excess beyond the disc edge and the deficit above 10\,TeV within $0.4^\circ$ of the source. Unlike the benchmark model, however, the field-trained model shows no gradient in Galactic latitude even for $\bm z_k=0$: having been trained on this field, it has learned the Galactic emission as part of the background. An in-field model thus treats all emission outside its training exclusions as background, which is convenient for the target but hides the neighboring emission from the analysis.

\begin{table}[h]
\centering
\footnotesize
\setlength{\tabcolsep}{2pt}
\begin{tabular}{llccccc}
\toprule
model & bkg. & $a$, $\sigma$ [$^\circ$] & $e$ & $\phi$ [$^\circ$] & $\Gamma$ & $N_0$ \\
\midrule
disc & BAccMod & 0.196(7) & 0.79(5) & 149(6) & 2.16(7) & 3.01(17) \\
disc & field & 0.203(2) & 0.77(1) & 147(2) & 2.27(7) & 3.27(15) \\
disc & benchmark & 0.204(10) & 0.78(3) & 145(5) & 2.23(7) & 3.20(22) \\
Gaussian & BAccMod & 0.107(8) & 0.78(6) & 145(8) & 2.15(7) & 3.08(17) \\
Gaussian & field & 0.115(8) & 0.76(7) & 144(8) & 2.27(6) & 3.46(18) \\
Gaussian & benchmark & 0.112(8) & 0.77(6) & 146(8) & 2.22(6) & 3.33(18) \\
\bottomrule
\end{tabular}
\caption{Classical 3D likelihood fits of MSH~15$-$52 between 0.5 and 10\,TeV with an elliptical uniform disc (semi-major axis $a$) or an elliptical Gaussian (width $\sigma$ along the major axis) and a power law ($N_0$ in $10^{-12}\,\mathrm{cm^{-2}\,s^{-1}\,TeV^{-1}}$ at 1.4\,TeV), with the in-sample BAccMod model and the field-trained and benchmark generative models, each scaled by a free normalization per run. The source centers agree to within $0.013^\circ$ in RA and $0.006^\circ$ in Dec.}
\label{tab:msh-infield-fit}
\end{table}

\end{document}